\documentclass[aps,pra,twocolumn,showpacs,superscriptaddress,a4paper,groupedaddress]{revtex4-1}
\usepackage{graphicx}    % needed for figures
\usepackage{dcolumn}    % needed for some tables
\usepackage{bm}             % for math
\usepackage{amssymb}   % for math
\usepackage{amsmath}    % for multiline equation
\usepackage{subfigure}    % for subfigure
\usepackage{braket}
\usepackage{mathrsfs}
\usepackage[export]{adjustbox}
\usepackage[mathscr]{euscript}
\usepackage[toc,page]{appendix}
\begin{document}

\title{Superradiance in Inverted Multi-level Atomic Clouds}
\author{R.T. Sutherland$^{1}$}
\email{rsutherl@purdue.edu}

\author{F. Robicheaux $^{1,2}$}
\email{robichf@purdue.edu}
\affiliation{$^{1}$Department of Physics and Astronomy, Purdue University, West Lafayette IN, 47907 USA}
\affiliation{$^{2}$Purdue Quantum Center, Purdue University, West Lafayette,
Indiana 47907, USA}

\date{\today}

\begin{abstract}
This work examines superradiance in initially inverted clouds of \textit{multi-level} atoms. We develop a set of equations that can approximately calculate the temporal evolution of $N$ coupled atoms. This allows us to simulate clouds containing hundreds of multi-level atoms while eschewing the assumption and/or approximation of symmetric dipole-dipole interactions. This treatment is used to explore the effects that dephasing caused by elastic dipole-dipole interactions, and competition between multiple transitions have on superradiance. Both of these mechanisms place strong parametrical restrictions on a given transition's ability to superradiate. These results are likely important to recent experiments that probe superradiance in Rydberg atoms.
\end{abstract}
\maketitle

\section{Introduction}\label{intro}
The fact that coherent radiation can dramatically alter physical phenomena \cite{dicke1954}  has resurfaced at the forefront of physics. Studies have shown that superradiance must be considered when studying atomic \cite{bettles2015,bettles2016, sutherland2016,sutherland2016_2,marek1979,scully2006,scully2007,kaiser2015,kaiser2008_2,bromley2016,walker2008,cote2007, grimes2016,kaiser2015,roof2016_2,crubellier1981,scully2009,scully2015, moi1983, ruostekoski2016,lee2016, jennewein2016}, biological \cite{yang2016, monshouwer1997_2}, and condensed matter systems \cite{bradac2016, scheibner2007, baumann2010}. Upon its recent revival, superradiance has led to conflicting ideas in the Rydberg atom community. At first glance, transitions between high-lying Rydberg states seem like perfect candidates for superradiance \cite{moi1983,cote2007,walker2008}. Because of their long wavelengths, Rydberg transitions allow experimentalists to reach the limit first described by Dicke \cite{dicke1954}, where atoms are placed in a volume with a radius that is very small relative to the wavelength. Later theoretical works \cite{friedberg1972, gross1982,moi1983}, however, argued that putting a cloud of atoms in this regime would result in large dipole-dipole interactions that quickly dephase the transition and destroy superradiance. This implies that transitions between high-lying Rydberg levels should \textit{not} superradiate. To exacerbate the confusion, conflicting experimental results have been reported, with some groups claiming to have either directly or indirectly observed Rydberg atom superradiance \cite{cote2007,walker2008,grimes2016,han2014}, and one claiming to have observed no superradiance at all \cite{zhou2016}. In order to unravel these experimental and theoretical differences, new developments are needed.

\begin{figure}[h]
	\includegraphics[width=0.45\textwidth]{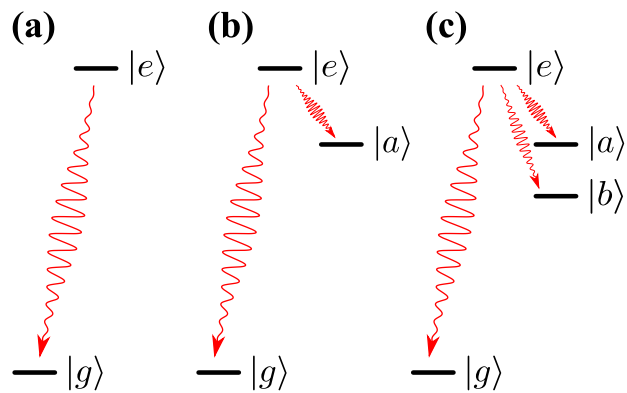}
	\caption{This paper studies superradiant cascades in clouds of (a) two, (b) three, and (c) four -level atoms. (a) Represents a two-level system coupled via dipole-dipole interactions. (b) Represents a three-level system, where dipole-dipole interactions for the $a$ transition are considered, but are not considered for the $g$ transition, due to the relatively small value of $\lambda_{g}$. (c) Represents a four-level system, where an additional interacting transition, $b$, has been added to the system represented by (b).} 
	\label{fig:pretty}
\end{figure}

The physics of superradiance in a vacuum is governed by the set of dipole-dipole interactions between every atom pair. These interactions can be traced back to two distinct types of photon exchanges, real and virtual. The exchange of real photons (see Eq.~(\ref{eq:real_photons})) contributes to the collective decay of the atomic ensemble. These \textit{inelastic} dipole-dipole interactions, result in either superradiance or subradiance. As the value of $|\vec{\boldsymbol{r}}_{i}-\vec{\boldsymbol{r}}_{j}|\rightarrow 0$, where $|\vec{\boldsymbol{r}}_{i} - \vec{\boldsymbol{r}}_{j}|$ is the spatial distance between atoms $i$ and $j$, the value of every inelastic coupling approaches $\Gamma_{\alpha}/2$, where $\Gamma_{\alpha}$ is the decay rate of the transition $e\rightarrow \alpha$ in an isolated atom (since the excited level is the same for every transition (see Fig. 1), the lower energy level, $\alpha$, is used as the label of a given transition). Resultantly, this part of the dipole-dipole interaction is completely symmetric for sufficiently dense clouds, i.e. the Dicke limit \cite{dicke1954}. Not included in Dicke's work are the interactions that result from exchanges of virtual photons (see Eq.~(\ref{eq:virual_photons})). These exchanges cause shifts between the energy levels of the system, often referred to as the collective/cooperative Lamb shifts \cite{friedberg1973}. These couplings will be referred to as \textit{elastic} dipole-dipole interactions from here on. In a vacuum, the magnitude of the elastic interaction between two atoms diverges $\propto 1/(k_{\alpha}|\vec{\boldsymbol{r}}_{i} - \vec{\boldsymbol{r}}_{j}|)^{3}$, where $k_{\alpha} \equiv 2\pi/\lambda_{\alpha}$. For dense atomic clouds, this results in large and random energy shifts between levels that dephase the system, quelling superradiance. This will be referred to as elastic dephasing in the rest of this paper.

A rigorous numerical treatment of this effect requires an implementation of the superradiance master equation \cite{gross1982}, or equivalently the quantum Monte-Carlo wave-function algorithm \cite{carmichael2000}. Unfortunately, both of these calculations grow exponentially with the number of atoms being simulated, $N$. So far, this has limited numerical calculations to systems such that $N \sim 10$ or less. While recent treatments have been developed that permit simulations involving large numbers of highly-excited atoms \cite{holland2013,hartmann2012,cote2007}, so far they rely on symmetries that result from either the assumption or approximation of symmetric dipole-dipole interactions. These approximations are not valid in this work because they ignore elastic dephasing.

Because of this, we derive and implement a numerical approach that scales $\propto N^{4}$, rather than exponentially. This enables the simulation of initially inverted clouds of hundreds of multi-level atoms. Note that a system containing initially inverted atoms is qualitatively different than a system where a particular transition is triggered, such as that in \cite{grimes2016}. Unlike previous methods, our approach can fully incorporate the inhomogeneous dipole-dipole interactions present in the cloud. This is done by solving the set of differential equations that describe the expectation values of the operators: $b^{\alpha -}_{i}b^{\alpha +}_{j}$, where $b^{\alpha -(+)}_{i}$ represents the lowering(raising) operator for the $e\rightarrow \alpha$ transition of the $i^{th}$ atom. The resulting equations are then truncated by factorizing the higher-order correlation operators. Since a full analysis of superradiating Rydberg atoms requires an understanding of the competition between many potentially superradiant transitions, the derivation assumes an initially excited state that can decay into an arbitrary number of lower energy states. The equations are then implemented in order to study the rich physics that occurs in clouds of two, three, and four -level atoms (see Fig.~\ref{fig:pretty}).

This paper is organized in the following way: first in Sec. \ref{sec:theory}, the equations that describe a system containing $N$ multi-level atoms are derived. In Sec. \ref{sec:two}, this system of equations is used to simulate a cloud of two-level atoms. Here, it is demonstrated how superradiance is limited by elastic dephasing in high density systems and by diffraction in low density systems. The result of this is that for a cloud with a given $N$ and density, $\mathcal{N}$, there is a particular value of $\lambda_{g}$, such that the coherent emission is at a maximum. In Sec. \ref{sec:three}, it is shown that when a system has multiple decay channels, superradiance develops in a very different manner than when there is only one. Here we mimic an elementary Rydberg system by including one transition to a high-lying state, $a$, with a value of $\lambda_{a}$ such that atoms couple via dipole-dipole interactions significantly. On top of this, we include one transition to the ground state, $g$, with a very small $\lambda_{g}$ such that dipole-dipole interactions are negligible (see Fig.~\ref{fig:pretty}(b)). This section shows that the presence of an alternate decay path strongly diminishes the buildup of superradiance. Finally, Sec. \ref{sec:four} shows the physics that results when multiple transitions can superradiate at once. This section indicates that the previously proposed mechanism for superradiance in Rydberg atoms \cite{cote2007}, where states tend to superradiate via transitions with the largest values of $\lambda_{\alpha}$, is likely not the dominant mechanism in many Rydberg atom systems. Here it is argued that only transitions with $\lambda_{\alpha}$ lying within a certain range of $\lambda_{\alpha}\mathcal{N}^{1/3}$ superradiate. This might lead to an explanation of the current experimental disagreements \cite{cote2007,zhou2016,walker2008,grimes2016}.

\section{Numerical Treatment}\label{sec:theory}
\subsection{Master Equation Evaluation}\label{sec:master}

The time dependence of the reduced density matrix, $\hat{\rho}$, is given by the master equation \cite{agarwal2012}

\begin{equation}\label{eq:master}
\frac{d{\hat{\rho}}}{dt} = -\frac{i}{\hbar}[H_{ed},\hat{\rho}] + \mathcal{L}\big(\hat{\rho}\big).
\end{equation}
Here $H_{ed}$ represents the elastic dipole-dipole interaction defined by the Hermitian Hamiltonian:

\begin{equation}\label{eq:h_dd}
H_{ed} = \sum_{i \neq j, \alpha}\hbar f^{\alpha}_{ij}b^{\alpha+}_{i}b^{\alpha-}_{j},
\end{equation}
where $b^{\alpha +}_i \equiv \ket{e_{i}}\bra{\alpha_{i}}$, $b^{\alpha -}_{i} \equiv \ket{\alpha_{i}}\bra{e_{i}}$, and
\begin{eqnarray}\label{eq:virual_photons}
f^{\alpha}_{ij} &=\nonumber& \frac{3\Gamma_{\alpha}}{4}\Big\{\Big(1 - 3\cos^{2}\phi^{\alpha}_{ij}\Big)\Big( \frac{\sin\xi^{\alpha}_{ij}}{\xi^{\alpha 2}_{ij}} + \frac{\cos\xi^{\alpha}_{ij}}{\xi^{\alpha 3}_{ij}}\Big) \\ &-&  \sin^{2}\phi^{\alpha}_{ij}\frac{\cos\xi^{\alpha}_{ij}}{\xi^{\alpha}_{ij}}\Big\}.
\end{eqnarray}
$\mathcal{L}(\hat{\rho})$ is the Lindblad superoperator given by:
\begin{equation}\label{eq:lindblad}
\mathcal{L}(\hat{\rho}) = \sum_{i,j,\alpha}\Gamma^{\alpha}_{ij}\Big(b^{\alpha-}_{j}\hat{\rho}b^{\alpha+}_{i} - \frac{1}{2}b^{\alpha+}_{i}b^{\alpha-}_{j}\hat{\rho} -\frac{1}{2}\hat{\rho}b^{\alpha+}_{i}b^{\alpha-}_{j} \Big),
\end{equation}
where $\Gamma^{\alpha}_{ij}$ is the inelastic dipole-dipole interaction of atoms $i$ and $j$ for the $e\rightarrow \alpha$ transition,

\begin{eqnarray}\label{eq:real_photons}
\Gamma^{\alpha}_{ij} &=\nonumber& \frac{3\Gamma_{\alpha}}{2}\Big\{ \Big(1 - 3\cos^{2}\phi^{\alpha}_{ij}\Big)\Big( \frac{\cos\xi^{\alpha}_{ij}}{\xi^{\alpha 2}_{ij}} - \frac{\sin\xi^{\alpha}_{ij}}{\xi^{\alpha 3}_{ij}} \Big) \\ &+& \sin^{2}\phi^{\alpha}_{ij}\frac{\sin\xi^{\alpha}_{ij}}{\xi^{\alpha}_{ij}} \Big\}.
\end{eqnarray}
In this notation, $\phi^{\alpha}_{ij}$ is the angle between the $\alpha^{th}$ dipole moment and the relative position of atoms $i$ and $j$, $\vec{\boldsymbol{r}}_{i} - \vec{\boldsymbol{r}}_{j}$. $\xi^{\alpha}_{ij} = k_{\alpha}|\vec{\boldsymbol{r}}_{i} - \vec{\boldsymbol{r}}_{j}|$, where $k_{\alpha}$ is the wavenumber of the $e\rightarrow \alpha$ transition, $2\pi/\lambda_{\alpha}$.

The fact that the system of interest starts in state $\ket{eee...e}$, and has no driving term, leads to a massive truncation of the Liouville-space. If one expresses $\hat{\rho}$ in the form:
\begin{equation}
\hat{\rho} = \sum_{m,n}c_{mn}\ket{m}\bra{n},
\end{equation}
it can be shown straightforwardly that the operators in Eq.~(\ref{eq:master}) will \textit{only} connect to elements of $\hat{\rho}$ such that $\ket{m}$ and $\bra{n}$ contain the same number of atoms in each level. This constitutes only a small fraction of $\hat{\rho}$. For a system of $N$ two-level atoms, the full Liouville-space contains $4^N$ matrix elements. However, only
\begin{equation}
\sum_{k=0}^{N}\binom{N}{k}^{2} = \binom{2N}{N}
\end{equation}
of these elements are actually non-zero. Incorporating this into our numerical algorithm, enables the exact calculation of Eq.~(\ref{eq:master}) up to $10$ atoms (see Fig. (\ref{fig:approx})).
\subsection{Approximate Evaluation of Operators}\label{sec:opterators}

The problems addressed in this paper require simulating hundreds of atoms, while avoiding the usual mean-field approximations that ultimately ignore elastic dephasing \cite{holland2013,hartmann2012,cote2007}. This is accomplished by solving the differential equations that describe the probability of atom $i$ being in state $\alpha$, or in operator form: $\langle b^{\alpha -}_{i}b^{\alpha +}_{i}\rangle$, as well as the quadratic correlation functions for atoms $i$ and $j$, defined as $\langle b^{\alpha -}_{i}b^{\alpha +}_{j} \rangle$. The change in the expectation value of an operator, $\Omega$, with time is determined by:

\begin{equation}\label{eq:one}
\frac{d}{dt}\langle\Omega\rangle = Tr\Big\{\Omega\frac{d{\hat{\rho}}}{dt}\Big\}.
\end{equation}
Using Eq.~(\ref{eq:master}) and Eq.~(\ref{eq:one}), one can derive $\frac{d}{dt}\langle b^{\alpha-}_{i}b^{\alpha+}_{i}\rangle$ (see appendix). This gives:
\begin{eqnarray}
\frac{d}{dt}\langle b^{\alpha-}_{i}b^{\alpha+}_{i}\rangle &=& \Gamma_{\alpha}\Big(1 -\sum_{\beta}\langle b^{\beta -}_{i}b^{\beta +}_{i} \rangle\Big)\nonumber \\ &+& 2\sum_{j \neq i}Re\Big(g^{\alpha}_{ij}\langle b^{\alpha-}_{j}b^{\alpha+}_{i}\rangle \Big),
\end{eqnarray}
where $g^{\alpha}_{ij}$ is the complex dipole-dipole interaction ($g^{\alpha}_{ij} \equiv if^{\alpha}_{ij} + \Gamma^{\alpha}_{ij}/2$). These equations are dependent on the quadratic correlation functions, which may be solved for in the same manner (see appendix). This yields a system of equations that is not closed, because solving for the quadratic correlation functions results in equations that depend on the quartic correlation functions. To avoid this, the system of equations is truncated by factorizing the quartic correlation functions in the following manner:

\begin{eqnarray}\label{eq:approx}
\langle b^{\alpha -}_{n}b^{\alpha +}_{j}b^{\beta -}_{m}b^{\beta +}_{m} \rangle &\simeq & \langle  b^{\alpha -}_{n}b^{\alpha +}_{j}\rangle \langle  b^{\beta -}_{m} b^{\beta +}_{m} \rangle \nonumber \\
\langle b^{\beta -}_{m}b^{\beta +}_{j} b^{\alpha -}_{n} b^{\alpha +}_{m} \rangle &\simeq & \langle b^{\beta -}_{m}b^{\beta +}_{j}\rangle \langle  b^{\alpha -}_{n}  b^{\alpha +}_{m} \rangle \nonumber \\
 \langle b^{\alpha -}_{m}b^{\alpha +}_{m} b^{\beta -}_{n} b^{\beta +}_{n} \rangle &\simeq & \langle b^{\alpha -}_{m}b^{\alpha +}_{m}\rangle \langle  b^{\beta -}_{n} b^{\beta +}_{n} \rangle \nonumber \\
\langle b^{\alpha -}_{n}b^{\alpha +}_{j}b^{\alpha -}_{m}b^{\alpha +}_{m} \rangle &\simeq & \langle  b^{\alpha -}_{n}b^{\alpha +}_{j}\rangle \langle  b^{\alpha -}_{m} b^{\alpha +}_{m} \rangle \nonumber \\
\langle b^{\alpha -}_{m}b^{\alpha +}_{m} b^{\alpha -}_{n} b^{\alpha +}_{n} \rangle &\simeq & \langle b^{\alpha -}_{m}b^{\alpha +}_{m}\rangle \langle  b^{\alpha -}_{n} b^{\alpha +}_{n} \rangle,
\end{eqnarray}
where $\alpha \neq \beta$. These factorizations are choosen with two `rules' in mind. First, the quartic operators must be grouped such that both the raising and lowering operators act on the same transition, in order to insure a closed system of equations. When this is satisfied, the quartic operator representing the population of state $\alpha$ for atom $m$ (i.e. $b^{\alpha -}_{m}b^{\alpha +}_{m}$) is factored out. This is done so that the factorized terms are smaller when little decay has occured, making our approximations accurate at early times. Implementing these approximations results in the following closed system of equations:
\begin{eqnarray}\label{eq:main}
&\frac{d}{dt}&\langle b^{\alpha -}_{n}b^{\alpha +}_{m} \rangle = - \langle b^{\alpha -}_{n}b^{\alpha +}_{m}  \rangle\sum_{\beta} \Gamma_{\beta} \nonumber \\ 
&+& \sum_{j \neq m,n} g^{\alpha *}_{jm}\langle b^{\alpha -}_{n}b^{\alpha +}_{j} \rangle\Big\{1 - \langle  b^{\alpha -}_{m} b^{\alpha +}_{m} \rangle
 - \sum_{\beta}\langle b^{\beta -}_{m}b^{\beta +}_{m} \rangle \Big\}\nonumber \\ 
  &+& \sum_{j \neq m,n} g^{\alpha}_{nj} \langle  b^{\alpha -}_{j} b^{\alpha +}_{m}\rangle \Big\{ 1 - \langle  b^{\alpha -}_{n} b^{\alpha +}_{n} \rangle
  - \sum_{\beta} \langle b^{\beta -}_{n}b^{\beta +}_{n} \rangle \Big\} \nonumber \\ 
  &-& \sum_{\beta \neq \alpha}\sum_{j \neq m,n}g^{\beta}_{nj}\langle b^{\alpha -}_{n}b^{\alpha +}_{m} \rangle \langle b^{\beta -}_{j}b^{\beta +}_{n} \rangle \nonumber \\ 
 &-& \sum_{\beta \neq \alpha}\sum_{j \neq m,n}g^{\beta *}_{jm}\langle b^{\alpha -}_{n} b^{\alpha +}_{m} \rangle\langle b^{\beta -}_{m} b^{\beta +}_{j}\rangle  \nonumber \\ 
  &+& 2Re\big\{g^{\alpha}_{nm}\big\} (1 - \sum_{\beta}\langle b^{\beta -}_{n}b^{\beta +}_{n}\rangle)(1 - \sum_{\beta}\langle b^{\beta -}_{m}b^{\beta +}_{m}\rangle) \nonumber \\  
  &-& g^{\alpha}_{nm}\langle  b^{\alpha -}_{n} b^{\alpha +}_{n} \rangle (1 - \sum_{\beta}\langle b^{\beta -}_{m}b^{\beta +}_{m}\rangle) \nonumber \\   
  &-& g^{\alpha *}_{nm} \langle b^{\alpha -}_{m}b^{\alpha +}_{m} \rangle(1 - \sum_{\beta}\langle b^{\beta -}_{n}b^{\beta +}_{n}\rangle).
\end{eqnarray}

\begin{figure}[h]
	\includegraphics[width=0.45\textwidth]{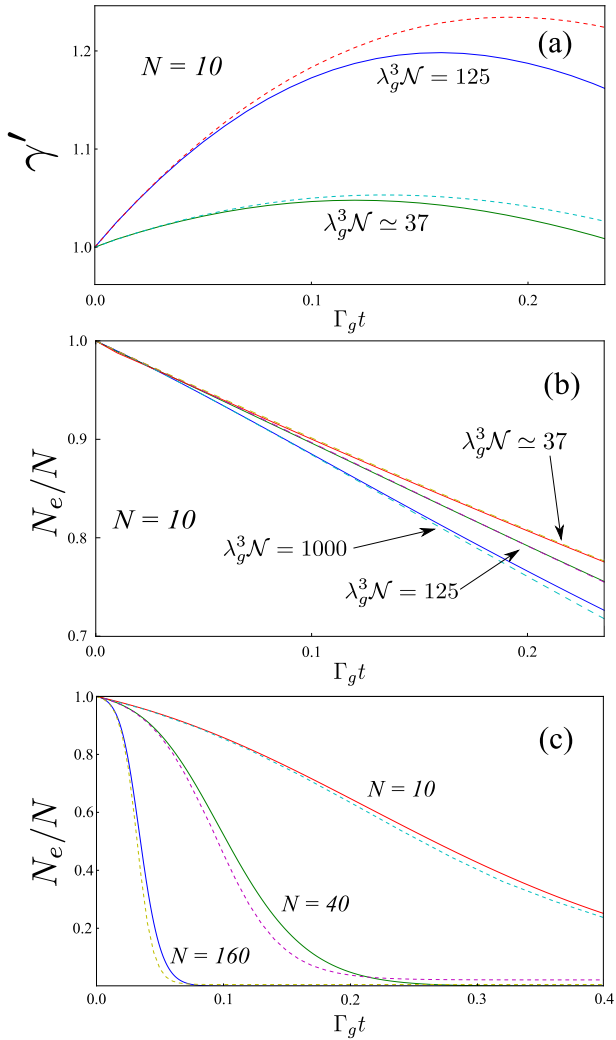}
	\caption{ Comparison of the time dependence resulting from Eq.~(\ref{eq:master}) (solid line) and Eq.~(\ref{eq:main}) (dashed line) for clouds of two-level atoms. (a) Photon emission rate per atom, $\gamma^{\prime}\equiv \gamma/N$, for clouds where $N=10$, and densities: $\lambda_{g}^{3}\mathcal{N} = 125$ and $\lambda_{g}^{3}\mathcal{N} \simeq 37$. (b) Excitation probability, $N_{e}/N$, for clouds where $N=10$, and $\lambda_{g}^{3}\mathcal{N} \simeq 1000, 125,$ and $37$. Note that in the figure, the values of $\lambda_{g}^{3}\mathcal{N}$ are inversly proportional to the slope of  $N_{e}$. (c) $N_{e}/N$ for $N = 10,~40$, and $160$ two-level atoms using the Dicke model, i.e. $\mathcal{N} \rightarrow \infty$ and $H_{ed} = 0$. Results are shown for the full calculation obtained using Eq.~(\ref{eq:master}), as well as the approximate Eq.~(\ref{eq:main}). Note that the two calculations scale with $N$ in the same manner.} 
	\label{fig:approx}
\end{figure}

The factorized terms are initially negligible. Therefore, at early times the results from Eq.~(\ref{eq:main}) agree quantitatively with the results from Eq.~(\ref{eq:master}). At later times, the two equations agree qualitatively. This is shown in Fig.~\ref{fig:approx}(a), where the photon emission rate per atom, $\gamma^{\prime}$, versus time is obtained by solving both Eq.~(\ref{eq:master}) and Eq.~(\ref{eq:main}) for clouds of 10 atoms. Fig.~\ref{fig:approx}(b) shows the probability of excitation:
\begin{equation}
\frac{N_{e}}{N} \equiv \frac{1}{N}\sum_{i}\langle b^{\alpha+}_{i}b^{\alpha -}_{i}\rangle
\end{equation}
versus time for $10$ atom clouds, using several values for density, $\mathcal{N}$. The figure not only shows how Eq.~(\ref{eq:main}) is qualitatively accurate at later times, but that it scales correctly with $\mathcal{N}$. Fig.~\ref{fig:v_dens} also shows this by demonstrating that the photon emission maxima versus $\mathcal{N}$ calculated with both methods, closely match for $10$ atom clouds.  Sec.~\ref{sec:size} further illustrates this fact for clouds where elastic dephasing is neglected (see Fig.~\ref{fig:v_dens_no_cls}). Figure \ref{fig:approx}(c) shows that Eq.~(\ref{eq:main}) also scales correctly with $N$. Here Eq.~(\ref{eq:main}) is compared with Eq.~(\ref{eq:master}) for the pure Dicke model (i.e. $\lambda_{g}^{3}\mathcal{N}\rightarrow \infty$, Im$(g_{ij}^{\alpha}) = 0$, and $H_{ed} = 0$) \cite{dicke1954}. Using this model and clouds such that $N = 10, 40,$ and $160$, the results from Eq.~(\ref{eq:main}) scale with $N$ in the same way as the results from Eq.~(\ref{eq:master}). Since the calculations of this work are intended to probe superradiance in regimes that are unreachable using Eq.~(\ref{eq:master}), the fact that the output of Eq.~(\ref{eq:main}) scales correctly with both $N$ and $\lambda_{g}^{3}\mathcal{N}$ indicates the validity of the calculations presented below.

One may note that the results of Fig.~\ref{fig:approx}(b) appear to agree significantly better than those of Fig.~\ref{fig:approx}(a). This is because the photon emission rates, shown in Fig.~\ref{fig:approx}(a), correspond to the negative derivatives of the calculations in Fig.~\ref{fig:approx}(b). These small differences in slope, produce very small differences in $N_{e}$ over the time frame shown. One may also note some differences in the calculations shown in Fig.~\ref{fig:approx}(c), at longer times. For example, in the simulation of a cloud containing $40$ atoms, the value of $N_{e}/N$ approaches approximately $0.022$ at later times, rather than $0$. This illustrates the fact that the results of Eq.~(\ref{eq:main}) are only exact when a small amount of population has decayed via an interacting transition. 

For simplicity, all of the states considered here are assumed to be $M_{J}=0$. Unless specified otherwise, all calculations average over $15360/N$ randomly generated frozen atomic clouds. Each cloud is given a Gaussian density distribution:
\begin{equation}\label{eq:gaussian}
\mathcal{N}(r) = \frac{N}{\sigma^{3}(2\pi)^{3/2}}\exp\Big(\frac{-r^{2}}{2\sigma^{2}}\Big),
\end{equation}
with an average density, $\mathcal{N}$, determined by $\mathcal{N} = N/(4\pi\sigma^{2})^{3/2}.$

\section{Two-Level Atoms}\label{sec:two}

The Dicke model describes two-level atoms (see Fig.~\ref{fig:pretty}(a)) radiating from a volume that is small relative to the transition's wavelength, $\lambda_{g}$. In this limit, all the atoms radiate from effectively the same position, making their emission in all directions coherent. This dissipative coherence causes the increase in photon emission rate associated with superradiance \cite{dicke1954}. In reality, the system is more complex in two important ways: experimentally realizable clouds can be much larger than $\lambda_{g}$, and physical clouds undergo elastic dephasing. 

\subsection{Finite Size Effects}\label{sec:size}
Large and dilute clouds (compared to the Dicke limit) can also superradiate \cite{rehler1971,feld1973}. While atoms in such clouds are usually separated by distances larger than $\lambda_{g}$, the emission of successive photons in a particular direction, $\hat{\boldsymbol{k}}_{g}$, projects the cloud onto a quantum state with a diffraction maxima, and therefore coherent radiation, in $\hat{\boldsymbol{k}}_{g}$. This causes superradiance \cite{dicke1954,rehler1971,carmichael2000}. Ignoring elastic dephasing and invoking a semiclassical approximation, the time-dependence of a given cloud can be shown to be \cite{rehler1971}:
\begin{equation}\label{eq:dif_sim}
\dot{N_{e}} = -\Gamma_{g}\Big\{N_{e} + \boldsymbol{\mu}(k_{g}\sigma)N_{e}N_{g}\Big\},
\end{equation}
where $k_{g}\equiv 2\pi/\lambda_{g}$, $N_{g}$ is the expectation value of number of atoms in the ground state ($N_{g} = N - N_{e}$), $\Gamma_{g}$ is the single-atom decay rate of the $g$ transition, and $\boldsymbol{\mu}(k_{g}\sigma)$ is a shape parameter, which for a cloud of $\hat{\boldsymbol{x}}$ polarized two-level atoms is given by:
\begin{equation}\label{eq:shape}
\boldsymbol{\mu}(k_{g}\sigma) = \frac{3}{8\pi N^{2}}\int d\Omega_{k} \Big\{1 - (\hat{\boldsymbol{k}}\cdot\hat{\boldsymbol{x}})^{2}\Big\} \sum_{m\neq n} e^{ik_{g}(\hat{\boldsymbol{k}} - {\hat{\boldsymbol{k}}}_{g})\cdot(\vec{\boldsymbol{r}}_{m}-\vec{\boldsymbol{r}}_{n})},
\end{equation}
where $\hat{\boldsymbol{k}}_{g}$ is the direction of the diffraction maxima of the radiation. This equation shows that $\boldsymbol{\mu}(k_{g}\sigma)$ is determined by the diffraction pattern of the cloud's emission \cite{rehler1971}. Solving this non-linear differential equation for the time dependence of the cloud's photon emission rate, $\gamma(t)$, yields:
\begin{equation}\label{eq:time_dep}
\gamma(t) = \frac{N\Gamma_{g}\Big(1 + N\boldsymbol{\mu}(k_{g}\sigma)\Big)^{2}e^{\Gamma_{g} t(1 + N\boldsymbol{\mu}(k_{g}\sigma))}}{\Big(N\boldsymbol{\mu}(k_{g}\sigma) + e^{\Gamma_{g} t(1 + N\boldsymbol{\mu}(k_{g}\sigma))} \Big)^{2}},
\end{equation}
which reaches a radiative maximum after a time delay, $t_{d}$, equal to:

\begin{equation}\label{eq:time_delay}
t_{d} = \frac{\ln\big(N\boldsymbol{\mu}(k_{g}\sigma)\big)}{\Gamma_{g}\big(1 + N\boldsymbol{\mu}(k_{g}\sigma)\big)}.
\end{equation}
This equation may be applied to the present system, a spherically symmetric Gaussian cloud, by converting the discrete sums in Eq.~(\ref{eq:shape}) to integrals, assuming $\hat{\boldsymbol{k}}_{g} = \hat{\boldsymbol{z}}$. Performing the integrals over both $\hat{\boldsymbol{k}}$ and the atomic positions yields:
\begin{eqnarray}
\boldsymbol{\mu}(k_{g}\sigma) &=&\nonumber \frac{3}{32k_{g}^{6}\sigma^{6}}\Big\{1 - 2k_{g}^{2}\sigma^{2} + 4k_{g}^{4}\sigma^{4} \\ &-& e^{-4k_{g}^{2}\sigma^{2}}\big(1 + 2k_{g}^{2}\sigma^{2} + 4k_{g}^{4}\sigma^{4}\big)\Big\}.
\end{eqnarray}
In the limit, $k_{g}\sigma \rightarrow \infty$, $\boldsymbol{\mu}(k_{g}\sigma) \rightarrow \frac{3}{8k_{g}^{2}\sigma^{2}}$. This agrees with recent results that show the collective enhancement to the decay rate of a large, singly-excited, and dilute atomic cloud is $\frac{3(N-1)\Gamma_{g}}{8k_{g}^{2}\sigma^{2}}$ \cite{sutherland2016}. Conversely, in the small cloud ($k_{g}\sigma \rightarrow 0$) limit, $\boldsymbol{\mu}({k_{g}\sigma})\rightarrow 1$, which reduces the more general Eq.~(\ref{eq:dif_sim}) to the equation derived by Dicke \cite{dicke1954}. 

\begin{figure}[h]
	\includegraphics[width=0.45\textwidth]{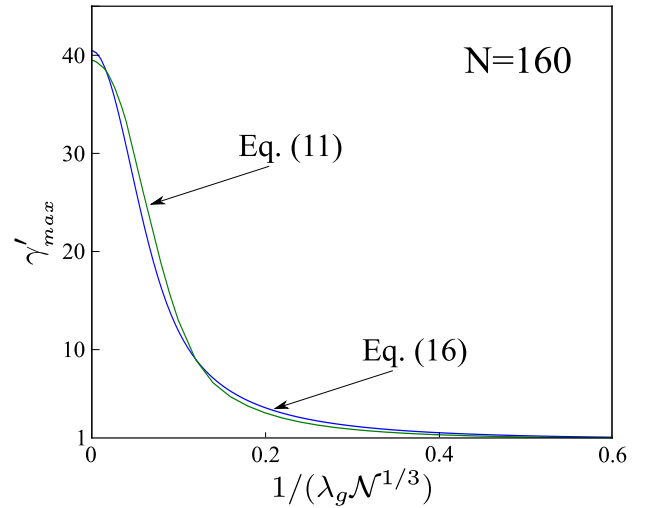}
	\caption{The maximum photon emission rate per atom, $\gamma^{\prime}_{_{max}}$, given by both Eq.~(\ref{eq:main}) and Eq.~(\ref{eq:time_dep}) for a Gaussian cloud of 160 atoms when elastic dephasing is neglected. Note that the plots agree well, indicating the validity of both equations in the absence of dephasing.}
	\label{fig:v_dens_no_cls}
\end{figure}

Eq.~(\ref{eq:time_dep}) is a reasonable approximation when dephasing due to elastic interactions can be ignored. This statement is made apparent by comparing Eq.~(\ref{eq:time_dep}) to Eq.~(\ref{eq:main}) when $g^{\alpha}_{mn} \rightarrow \Gamma^{\alpha}_{mn}/2$, thus artificially keeping only the inelastic part of the each dipole-dipole interaction. Since one of the key features of an initially inverted superradiating cloud is an increase in photon emission, the maximum photon emission rate, $\gamma(t_{d})$, is used to quantify the `amount' of superradiance in a given cloud. Figure \ref{fig:v_dens_no_cls} compares the two calculations by showing the maximum photon emission rate per atom, $\gamma^{\prime}_{_{max}} \equiv \gamma(t_{d})/(N\Gamma_{g})$, versus $1/(\lambda_{g}\mathcal{N}^{1/3})$ when $N=160$. As the value of $\lambda_{g}^{3}\mathcal{N}$ increases, $\gamma^{\prime}_{_{max}}$ increases as well. This increase of $\gamma^{\prime}_{_{max}}$ with $\mathcal{N}$ may be understood by considering the diffraction pattern of a cloud that has emitted a series of photons in a particular direction, $\hat{\boldsymbol{k}}$. As the value of $\lambda_{g}^{3}\mathcal{N}$ increases, the size of the atomic cloud decreases. The result of this is that $\boldsymbol{\mu}(k_{g}\sigma)$ increases due to the broadening of the diffraction maxima centered at $\hat{\boldsymbol{k}}$. As the size of the cloud decreases, photons can radiate coherently into more solid angles, until finally the Dicke limit is reached and photons are coherent in all directions ($\boldsymbol{\mu}(k_{g}\sigma)\rightarrow 1$). This is seen for both calculations in Fig.~\ref{fig:v_dens_no_cls}, when the value of $\gamma^{\prime}_{_{max}}$ increases towards a constant, as $1/(\lambda_{g}\mathcal{N}^{1/3})$ decreases towards $0$. The quantitative agreement shown in the two simulations is a good indication of the usefulness of Eq.~(\ref{eq:time_dep}) when elastic interactions are neglected. However, this is often not valid.

\subsection{Dephasing Due to Elastic Dipole-dipole Interactions}\label{sec:lamb_deph}

The seminal work of Dicke ignores the off-resonant, virtual photon exchanges that lead to elastic dipole-dipole interactions. In dense clouds, such as the ones described by Dicke, elastic interactions cause large and random energy shifts and can lead the system to dephase on timescales much shorter than its collective decay rate \cite{gross1982, friedberg1972}. This has been described semi-classically \cite{friedberg1972}, as well as numerically for small values of $N$ (i.e. N $= 3-10$) \cite{damanet2016,carmichael2000}. In this section, Eq.~(\ref{eq:main}) is used to simulate clouds containing up to $640$ initially inverted atoms. This enables the exploration of the fundamental limits that elastic dipole-dipole interactions have on superradiance. For a given atomic cloud, the calculations presented in this section place strong restrictions on the parameters that can lead to significant superradiant buildup. As will be shown in Sec. \ref{sec:four}, this has important implications on the Rydberg atom problem.

\begin{figure}[h]
	\includegraphics[width=0.45\textwidth]{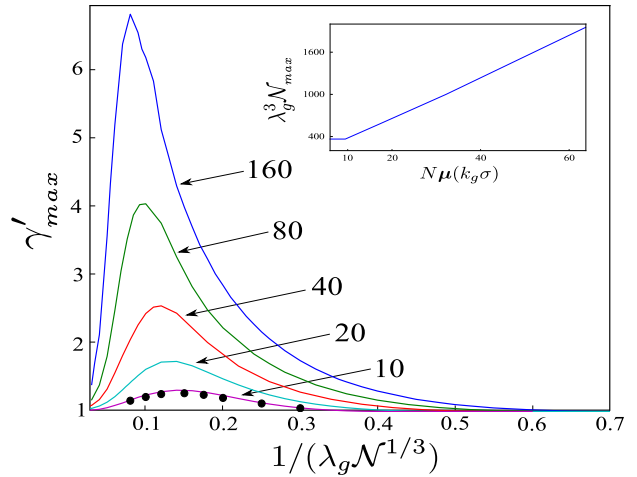}
	\caption{Eq.~(\ref{eq:main}) is used to calculate the maximum photon emission rate divided by the number of atoms, $N$, times the single atom decay rate, $\Gamma_{g}$, ($\gamma^{\prime}_{_{max}} \equiv \gamma(t_{d})/(N\Gamma_{g})$)  versus $1/(\lambda_{g}\mathcal{N}^{1/3})$. This is done for clouds such that $N = 10,~20,~40,~80,~160$. As $1/(\lambda_{g}\mathcal{N}^{1/3})$ decreases, at first $\gamma^{\prime}_{_{max}}$ increases, due to the broadening of the diffraction maxima. However, for every value of $N$ there is a certain $\lambda_{g}^{3}\mathcal{N}$ where the large elastic dipole-dipole interactions begin to dominate, and the value of $\gamma^{\prime}_{_{max}}$ starts to decrease rapidly. The inset shows that the value of $\lambda_{g}^{3}\mathcal{N}$ such that superradiance is at a maximum , $\lambda_{g}^{3}\mathcal{N}_{_{max}}$, increases linearly with $N\boldsymbol{\mu}(k_{g}\sigma)$ due to the collective enhancement to the decay rate. Note that in the main figure, the solid black dots represent the calculations for $10$ atom clouds using Eq.~(\ref{eq:master}).}
	\label{fig:v_dens}
\end{figure}

The relevant quantity when determining the dephasing rate of a specific atomic cloud is $\lambda_{g}^{3}\mathcal{N}$. In Fig. \ref{fig:v_dens}, Eq.~(\ref{eq:main}) is solved in order to demonstrate how $\gamma^{\prime}_{_{max}}$ for a particular transition depends on $N$ and $1/(\mathcal{N}^{1/3}\lambda_{g})$. In Fig.~\ref{fig:v_dens}, $\gamma^{\prime}_{_{max}}$ is shown for clouds such that $N = 10,~20,~40,~80,$ and $160$. There are several important effects visible here. First, increasing $\mathcal{N}$ in dilute clouds causes an increase in $\gamma^{\prime}_{_{max}}$. This is because in dilute clouds elastic dephasing is slow relative to the collective decay of the cloud. Therefore little dephasing occurs within $t_{d}$. Thus decreasing the cloud size simply increases the directional coherence (the value of $\boldsymbol{\mu}(k_{g}\sigma)$) discussed in Sec.~\ref{sec:size}. As one increases $\lambda_{g}^{3}\mathcal{N}$, however, the dephasing rate due to elastic interactions (Eq.~(\ref{eq:virual_photons})) grows linearly, while the radiative enhancement due to inelastic interactions (Eq.~(\ref{eq:real_photons})) approaches a constant value ($\Gamma_{g}/2$). Therefore, in dense clouds the system dephases significantly before $t_{d}$, and the value of $\gamma^{\prime}_{_{max}}$ begins to diminish $\propto\lambda_{g}^{3}\mathcal{N}$. This is seen in Fig.~\ref{fig:v_dens}, where $\gamma^{\prime}_{_{max}}$ increases as $1/(\lambda_{g}\mathcal{N}^{1/3})$ decreases for dilute clouds, followed by a rapid decrease as $1/(\lambda_{g}\mathcal{N}^{1/3})\rightarrow 0$. Due to computational limitations, the largest value of $\lambda^{3}_{g}\mathcal{N}$ in Fig.~\ref{fig:v_dens} is $37037$, however, the values of $\gamma^{\prime}_{_{max}}$ do seem to approach $0$ for increasingly dense clouds. This is in contrast to Fig.~\ref{fig:v_dens_no_cls}, where there is no elastic dephasing. Here, as clouds condense the value of $\gamma^{\prime}_{_{max}}$ increases to a constant with approximately 6 times the maximum values of $\gamma^{\prime}_{_{max}}$ seen in Fig.~\ref{fig:v_dens}. In order to demonstrate the accuracy of our approximate simulations, Fig.~\ref{fig:v_dens} compares the photon emission maxima for $10$ atom clouds using both Eq.~(\ref{eq:master}) and Eq.~(\ref{eq:main}). Note that for a given density, there is a narrow range of $\lambda_{g}$ where superradiance maximizes. 

Figure \ref{fig:v_dens} shows that for clouds with more atoms, the value of $\lambda_{g}^{3}\mathcal{N}$ with the largest value of $\gamma_{_{max}}^{\prime}$, $\lambda_{g}^{3}\mathcal{N}_{_{max}}$, increases. This is because the collective decay rate of a cloud increases $\propto N\boldsymbol{\mu}(k_{g}\sigma)$ while the dephasing rate of a dense cloud increases $\propto \lambda_{g}^{3}\mathcal{N}$. Assuming $\lambda_{g}^{3}\mathcal{N}_{_{max}}$ occurs when the two rates are equal, up to some constant, $\lambda_{g}^{3}\mathcal{N}_{_{max}}$ should increase linearly with $N\boldsymbol{\mu}(k_{g}\sigma)$. This increase of $\mathcal{N}_{_{max}}$ with $N\boldsymbol{\mu}(k_{g}\sigma)$ is shown in the inset of Fig. (\ref{fig:v_dens}).

\begin{figure}[h]
	\includegraphics[width=0.45\textwidth]{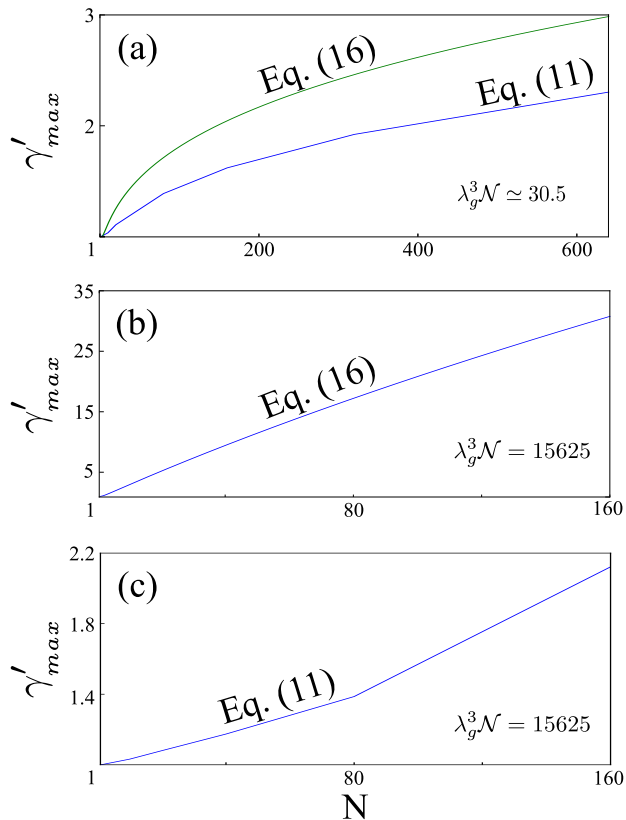}
	\caption{For clouds with a given density, $\mathcal{N}$, this figure shows the maximum photon emission rate per atom, $\gamma^{\prime}_{_{max}}$, versus the number of atoms, $N$. (a) For dilute clouds, where elastic dipole-dipole interactions may be ignored, the analytic results of Eq.~(\ref{eq:time_dep}) are similar to the full numerical treatment of Eq.~(\ref{eq:main}). Here a cloud with $\mathcal{N}=30.5/\lambda_{g}^{3}$ is shown. Note that for both calculations, the slope of $\gamma^{\prime}_{_{max}}$ versus $N$ decreases at larger $N$. This is due to the diffractive coherence as described in the text.  For more dense clouds where $\mathcal{N} = 15625/\lambda_{g}^{3}$, (b) shows the analytic results given by Eq.~(\ref{eq:time_dep}), and (c) shows the full numerical results obtained by solving Eq.~(\ref{eq:main}). Here the presence of large elastic dipole-dipole interactions in (c) causes the two calculations to notably deviate.
	}
	\label{fig:v_n}
\end{figure}

In the single excitation regime, it has been demonstrated that large elastic interactions produce negligible effects when clouds are sufficiently dilute. This enables relatively straightforward analytic treatments that agree well with the exact numerical results \cite{sutherland2016,bromley2016, ruostekoski2016, pellegrino2014}. For more dense clouds, however, the numerical results begin to deviate from the analytic ones \cite{bromley2016,sutherland2016}. Figure \ref{fig:v_n} shows that clouds of initially inverted atoms are similar in this respect. As seen in Fig.~\ref{fig:v_n}(a), for initially inverted and dilute clouds, Eq.~(\ref{eq:time_dep}) gives similar results to the full calculation of Eq.~(\ref{eq:main}). For lower values of $\lambda_{g}^{3}\mathcal{N}$, the numerical calculation of $\gamma^{\prime}_{_{max}}$ grows with $N$ in a similar manner to Eq.~(\ref{eq:time_dep}). For more dense clouds, the results are very different. When comparing Fig~\ref{fig:v_n}(b) and \ref{fig:v_n}(c), we see that the presence of large elastic dipole-dipole interactions significantly decreases the rate at which $\gamma^{\prime}_{_{max}}$ grows with $N$. Counterintuitively, in clouds where elastic dephasing is important, such as  in Fig.~\ref{fig:v_n}(c), the slope of $\gamma^{\prime}_{_{max}}$ increases with $N$. This is because the larger the value of $N\boldsymbol{\mu}(k_{g}\sigma)$, the sooner $t_{d}$ occurs, allowing a cloud less time to dephase before it decays significantly.

\subsection{Parallel with Classically Radiating Dipoles}
There is a notable parallel between the physics of a cloud of two-level atoms and that of a cloud of coupled harmonic oscillators. In matrix form, the set of equations that describes coupled harmonic oscillators is given by:
\begin{equation}
\dot{\vec{\boldsymbol{a}}} = -\underline{G}\vec{\boldsymbol{a}},
\end{equation}
where each element of the vector $\vec{\boldsymbol{a}}$ represents the complex amplitude of a specific oscillator. The matrix elements of the complex symmetric matrix, $\underline{G}$, are given by:
\begin{equation}
G_{mn} \equiv \frac{\Gamma^{g}_{mn}}{2} + if^{g}_{mn}(1 - \delta_{mn}).
\end{equation}
The eigenvalues of \underline{G} often have important physical significance \cite{goetschy2011,sutherland2016_2,bienaime2012,kaiser2015,ruoskekoski2016}, where the real part of an eigenvalue corresponds to half of that eigenmode's decay rate and the imaginary part corresponds to its energy shift, often called the collective/cooperative Lamb shift. This section will be concerned with each eigenmode's decay rate, $\Gamma_{j}$, defined by the equation:
\begin{equation}
\underline{G}\vec{\boldsymbol{a}}_{j} = \Big(\frac{\Gamma_{j}}{2} + i\epsilon_{j}\Big)\vec{\boldsymbol{a}}_{j}.
\end{equation}
In the original Dicke model, when a cloud decays, it cascades through a set of states with a specific `cooperation number'. For an initially inverted cloud of atoms, the cloud cascades through the most superradiant states (cooperation number = $N/2$) until it reaches the ground state \cite{dicke1954}. Since the systems investigated in this paper are initially inverted, we imagine a physical situation where the atomic cloud in question has decayed into the `most' superradiant  singly excited eigenmode, as a parallel to the results of Fig.~\ref{fig:v_dens}. 

\begin{figure}[h]
	\includegraphics[width=0.45\textwidth]{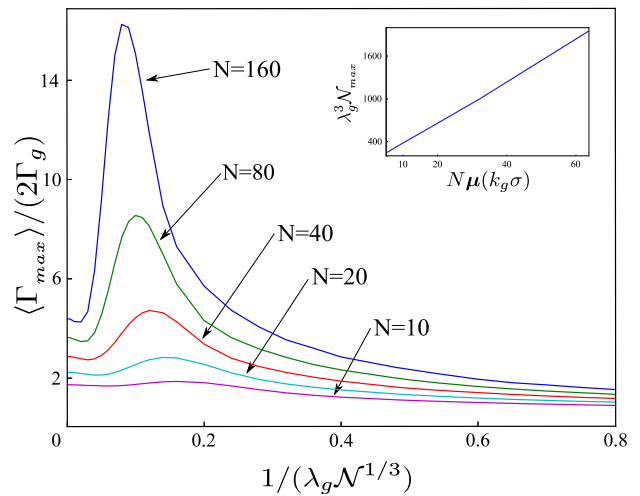}
	\caption{The average value of the decay rate of the most superradiant eigenmode, $\langle \Gamma_{_{max}} \rangle$, in clouds such that $N = 10, 20, 40, 80,$ and $160$. The inset shows the value of $\lambda_{g}^{3}\mathcal{N}$ such that $\langle \Gamma_{_{max}} \rangle$ is at a maximum, $\lambda_{g}^{3}\mathcal{N}_{_{max}}$.}
	\label{fig:eigen}
\end{figure}

In Fig.~\ref{fig:eigen}, the maximum value of $\Gamma_{j}$, $\langle \Gamma_{_{max}}\rangle$, averaged over $1.6\times 10^{5}/N$ runs, is shown for clouds where $N = 10, 20, 40, 80$, and $160$. More configurations are averaged here than in the rest of the paper, since only one data point from each run is kept. Fig.~\ref{fig:eigen} shows a pattern that is remarkably similar to Fig.~\ref{fig:v_dens}. At low densities, $\langle \Gamma_{_{max}}\rangle$ increases with $\lambda_{g}^{3}\mathcal{N}$ followed by a sharp decrease as the clouds become condensed. Also, as seen in the inset of Fig.~\ref{fig:eigen}, for every cloud with more than 10 atoms, the value of $\lambda_{g}^{3}\mathcal{N}_{_{max}}$ is equivalent to its corresponding value in the inset of Fig.~\ref{fig:v_dens}. Similar to a cascade of initially inverted atoms, there is competition between the increasingly symmetric inelastic interactions (Eq.~(\ref{eq:real_photons})) in $\underline{G}$, and the highly disordered and increasingly large elastic dipole-dipole interactions (Eq.~(\ref{eq:virual_photons})). Note that the values of $\langle \Gamma_{_{max}} \rangle$ in Fig.~\ref{fig:eigen} do not approach $0$ as $1/(\lambda_{g}\mathcal{N}^{1/3})$ approaches $0$. This is likely due to the fact that the eigenmodes in highly dense clouds become localized over several atoms \cite{sun2008}.

\section{Three-Level Atoms: Multiple Decay Channels}\label{sec:three}

Two-level systems, in many respects, are the best possible scenario for maximizing the effects of superradiance. Realistically, superradiance experiments typically involve populating an excited state that can decay via multiple transitions. These transitions then `compete' for the buildup in cooperativity that results in superradiance. As an example, a \textit{Rydberg-like} system (see Fig.~\ref{fig:pretty}(b)), of an exited state that can decay into a high-lying Rydberg state, $a$, as well as a low-lying ground state, $g$, is considered. For typical Rydberg systems $\lambda_{a} \gg \lambda_{g}$. Since Rydberg experiments are usually conducted for values of $\mathcal{N}$ such that $\lambda_{g}^{3}\mathcal{N} \ll 1$, setting $\Gamma^{g}_{ij}= 0$ and $f^{g}_{ij} = 0$ is valid. This is \textit{not} the case for transition $a$, where in many experimental setups $\lambda_{a}^{3}\mathcal{N} \gg 1$. Despite this potential for cooperative behavior, the presence of an alternate decay route significantly dampens the buildup of superradiance in the $a$ transition.

The qualitative physics studied here can be gleaned by examining the two linearly independent equations that correspond to Eq.~(\ref{eq:dif_sim}) for the three-level system described here:
\begin{eqnarray}
\dot{N_{e}} &=\nonumber & -(\Gamma_{g} + \Gamma_{a})N_{e} - \Gamma_{a}\boldsymbol{\mu}(k_{a}\sigma)N_{e}N_{a} \\
\dot{N_{a}} &=& \Gamma_{a}N_{e} + \Gamma_{a}\boldsymbol{\mu}(k_{a}\sigma)N_{e}N_{a},\label{eq:3}
\end{eqnarray}
where $k_{a} \equiv 2\pi/\lambda_{a}$. Note that $N = N_{g} + N_{a} + N_{e}$ and $0 = \dot{N_{g}} + \dot{N_{a}} + \dot{N_{e}}$. Since transition $a$ can potentially superradiate, one may imagine that if $N\boldsymbol{\mu}(k_{a}\sigma)\Gamma_{a} > \Gamma_{g}$, then the $a$ transition should dominate the $g$ transition. However, this is often not the case since the superradiant enhancement to $\dot{N_{a}}$ is proportional to the number of decays via transition $a$ ($N_{a}$). In Rydberg systems, $\Gamma_{a}/\Gamma_{g}$ is a small value. Therefore, until $\Gamma_{a}\boldsymbol{\mu}(k_{a}\sigma)N_{a} \sim \Gamma_{g}$, $\dot{N_{a}}$ will be smaller than $\dot{N_{g}}$. Often the entire system will decay before this occurs, preventing any significant superradiant enhancement to the transition $a$. 

\begin{figure}[h]
	\includegraphics[width=0.45\textwidth]{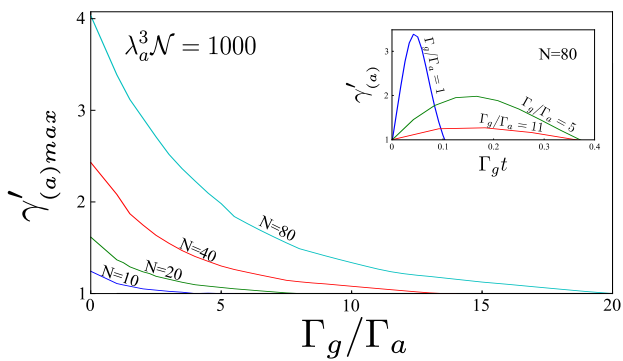}
	\caption{Eq.~(\ref{eq:main}) is used to calculate the maximum photon emission rate per atom for transition $a$ divided by $\Gamma_{a}$, $\gamma^{\prime}_{_{(a)max}}$, versus $\Gamma_{g}/\Gamma_{a}$. This is shown for clouds such that $\mathcal{N} = 1000/\lambda_{a}^{3}$ and  $N = 10, 20, 40,$ and $80$. For every value of $N$ shown, $\gamma^{\prime}_{_{(a)max}}$ decreases as the relative decay rate into state $g$ increases. The inset shows the temporal dependence of $\gamma^{\prime}_{_{(a)}}/\Gamma_{a}$. Note that when $\Gamma_{g}/\Gamma_{a}$ increases, $t_{d}$ increases.}
	\label{fig:three}
\end{figure}

The superradiant enhancement to the decay rate is illustrated through value of $\gamma^{\prime}$ for transition $a$, $\gamma^{\prime}_{_{(a)}}$. Fig.~\ref{fig:three} shows $\gamma_{_{(a)max}}^{\prime}$ versus $\Gamma_{g}/\Gamma_{a}$ calculated using Eq.~(\ref{eq:main}). Here the presence of an alternate decay channel quells superradiant behavior. This is seen in the fact that when the value of $\Gamma_{g}/\Gamma_{a}$ is increased, a decrease in $\gamma^{\prime}_{_{(a)max}}$ follows. As shown in Fig.~\ref{fig:three}, for clouds with larger values of $N$, the system is progressively more resistant to decay into state $g$. This is because clouds consisting of more atoms must radiate more photons before all atoms reach the ground state. This provides more opportunities for the system to decay into state $a$. When the system has decayed sufficiently into state $a$, such that $N_{a}\boldsymbol{\mu}(k_{a}\sigma)\Gamma_{a} \gg \Gamma_{g}$, the presence of transition $g$ becomes unimportant. This means that alternate decay paths are less relevant for clouds with more atoms. However, in many Rydberg-atom schemes $\Gamma_{g}/\Gamma_{a} \sim 50$, making this mechanism likely to be at least quantitatively important. Since the much simpler Eq.~(\ref{eq:3}) includes the above physics, the qualitative features of Fig.~\ref{fig:three} can be replicated using this equation; however, elastic dephasing dampens the results by a factor of approximately 2. We also note that the larger the value of $\Gamma_{g}/\Gamma_{a}$ is, the more accurate Eq.~(\ref{eq:main}) becomes. This is because the terms that are approximately factorized in Eq. (\ref{eq:main}) are smaller when less population decays via channel a.

As far as the superradiant enhancement to transition $a$ is concerned, the process of atoms decaying via an alternate decay route is similar to removing atoms from the system. Resultantly, the temporal dependence of the transition $a$ for larger values of $\Gamma_{g}/\Gamma_{a}$ is like that of a two-level system with less atoms. This can be seen in the inset of Fig.~\ref{fig:three}. Note that similar to clouds of two-level atoms (see Eq. (\ref{eq:time_delay})), $t_{d}$ 
increases for larger values of $\Gamma_{g}/\Gamma_{a}$.

\section{Four-Level Atoms: Competing Superradiant Channels}\label{sec:four}

In Rydberg atom experiments, there are multiple potentially superradiant transitions that compete with each other. To properly explore this, an additional transition to an upper-lying Rydberg state, $b$, is added to the system of the previous section (see Fig.~\ref{fig:pretty}(c)). This allows simulations of more realistic Rydberg systems. Here, the development of superradiance in a particular transition not only competes with the decay to state $g$, as described in the previous section, but also with another superradiating channel. For an isolated atom, the competition between multiple transitions can be summarized, straightforwardly, by the branching ratios of those transitions. These branching ratios are determined by the single atom decay rates of the system. When an ensemble of atoms radiates coherently, however, the competition between transitions is more complex. As will be shown, when considering the competition between superradiating transitions in a cloud, there are two important parameters: the single atom decay rates, and the relative densities for each transition (i.e. $\lambda^{3}_{\alpha}\mathcal{N}$).

If elastic dephasing is neglected, the superradiant enhancement to a transition is proportional to the number of atoms that have decayed via that transition (see Eq.~(\ref{eq:dif_sim})). It follows that transitions with larger decay rates will experience more coherent enhancement, simply because they decay more. In Sec. \ref{sec:two}, it was demonstrated that for a given value of $\mathcal{N}^{1/3}$ there is a very narrow range in $\lambda_{\alpha}$ such that transition $\alpha$ can develop strong superradiant character. This is because the superradiant enhancement to transition $\alpha$  is limited by diffraction if $\lambda_{\alpha}\mathcal{N}^{1/3}$ is too small and by elastic dephasing if $\lambda_{\alpha}\mathcal{N}^{1/3}$ is too large. Resultantly, a given Rydberg state will have only a small number of transitions with the potential to develop significant superradiant behavior. This differs from previous treatments ignoring elastic dephasing, since they argue that transitions with the largest values of $\lambda_{\alpha}$ always show the most superradiant enhancement \cite{rehler1971,cote2007}.

\begin{figure}[h]
	\includegraphics[width=0.45\textwidth]{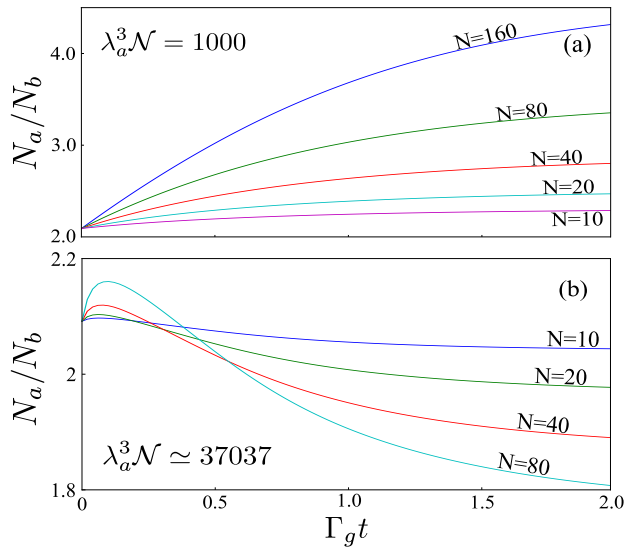}
	\caption{An atomic cloud is placed in a Rydberg state, $e$, that subsequently decays into three states: $a, b, $ and $g$. The values of $\Gamma_{\alpha}$ and $\lambda_{\alpha}$ for each transition are given in the text. Only transitions to states $a$ and $b$ experience dipole-dipole interactions.  For both of the figures above, the competition between the two transitions' superradiant behavior is illustrated by the ratio of the population in states $a$ and $b$, $N_{a}/N_{b}$. This is calculated using Eq.~(\ref{eq:main}). (a) $N_{a}/N_{b}$ versus time for clouds such that $\mathcal{N} = 1000/\lambda_{a}^{3}$. For more dilute clouds, the dephasing rate for both transitions is relatively low, causing the enhancement to the $a$ transition's decay rate to grow with $N$ faster than the $b$ transition's enhancement. (b) $N_{a}/N_{b}$ versus time for $\mathcal{N} \simeq 37037/\lambda_{a}^{3}$. Initially, the coherent enhancement to the $a$ transition's decay rate is stronger than the $b$ transition's enhancement. However, the elastic dipole-dipole interactions are much larger for the $a$ transition than for the $b$ transition. This causes the $a$ transition to quickly dephase, while the coherent enhancement to the $b$ transition's decay rate continues to build with $N$.}
	\label{fig:four}
\end{figure}

These effects are made apparent in Fig.~\ref{fig:four}. So as to obtain experimentally relevant results, Eq.~(\ref{eq:main}) is solved using values of $\Gamma_{\alpha}$ and $\lambda_{\alpha}$ calculated for specific transitions of Rb. For the calculations of Fig. \ref{fig:four}, $\Gamma_{a} = 169$ $s^{-1}$ and $\lambda_{a} = 1.134\times 10^{-3}$ m corresponding to the $26p\rightarrow 26s$ transition, while $\Gamma_{b} = 80.8$ $s^{-1}$ and $\lambda_{b} = 3.51\times 10^{-4}$ m corresponding to the $26p \rightarrow 25s$ transition. Lastly, the numbers for transition $g$ correspond to the $26p\rightarrow 5s$ transition, where $\Gamma_{g} = 3.5\times 10^{3}$ $s^{-1}$ and $\lambda_{g}$ is negligibly small. 

In Fig.~\ref{fig:four}, the value of $N_{a}/N_{b}$ versus time allows one to observe the temporal behavior of the two transitions' collective decay rates. Initially, $N_{a}/N_{b} = \Gamma_{a}/\Gamma_{b}$ because superradiant behavior has not developed yet. As the system begins to decay however, this ratio tends towards the transition experiencing the most collective enhancement. For example, Fig.~\ref{fig:four}(a) shows that for lower density clouds, $\mathcal{N} = 1000/\lambda_{a}^{3} \simeq 29.65/\lambda_{b}^{3}$, superradiant behavior develops the most in transition $a$. This occurs in part because $\Gamma_{a} > \Gamma_{b}$. Also, since $\lambda_{a}^{3}\mathcal{N}$ is not large enough for elastic dephasing to be important, the fact that $\boldsymbol{\mu}(k_{a}\sigma) > \boldsymbol{\mu}(k_{b}\sigma)$ causes the superradiant development to favor transition $a$. Therefore, for more dilute clouds, as $N$ is increased, the system tends to decay more into state $a$. This indicates that at this density, for clouds with $N \sim 10^{4}$, such as those of recent experiments \cite{cote2007,walker2008,zhou2016}, the $a$ transition is likely to dominate.

Conversely, Fig.~\ref{fig:four}(b) shows that for more condensed clouds, $\mathcal{N} \simeq 37037/\lambda_{a}^{3} \simeq 1098.3/\lambda_{b}^{3}$, elastic dephasing can be fast enough to significantly diminish superradiance in one transition, while still allowing superradiance to build in another. Here we can see that even though $\Gamma_{a} > \Gamma_{b}$ and $\boldsymbol{\mu}(k_{a}\sigma) > \boldsymbol{\mu}(k_{b}\sigma)$, at longer times $N_{a}/N_{b}$ decreases with $N$. Earlier in the evolution, $N_{a}/N_{b}$ in Fig. \ref{fig:four}(b) grows in a similar manner as Fig.~\ref{fig:four}(a) both because $\Gamma_{a} > \Gamma_{b}$ and because $\boldsymbol{\mu}(k_{a}\sigma) > \boldsymbol{\mu}(k_{b}\sigma)$. However, the elastic dephasing rate is significantly larger for transition $a$ than for transition $b$. Resultantly, later in the cloud's evolution, elastic dephasing causes the buildup of superradiance in transition $a$ to diminish, which in turn allows the collective behavior to favor transition $b$.

It is probable that for clouds with enough atoms, transition $a$ will likely dominate again. This is because, as is shown in the inset of Fig.~\ref{fig:v_dens}, the value of $\lambda^{3}_{a}\mathcal{N}_{_{max}}$ increases with $N$. If a given cloud contained enough atoms such that $\lambda^{3}_{a}\mathcal{N}_{_{max}} \simeq 37037$, then transition $a$ would likely re-emerge as the dominant superradiating transition at all times. Traces of this are visible in Fig. \ref{fig:four}(b), where for clouds with larger values of $N$, $N_{a}/N_{b}$ increases for slightly longer periods of time before elastic dephasing decimates it. If a cloud reaches the point where the entire system decays before elastic dipole-dipole interactions can significantly dephase transition $a$, then it will likely dominate the cascade.

\begin{figure}[h]
	\includegraphics[width=0.45\textwidth]{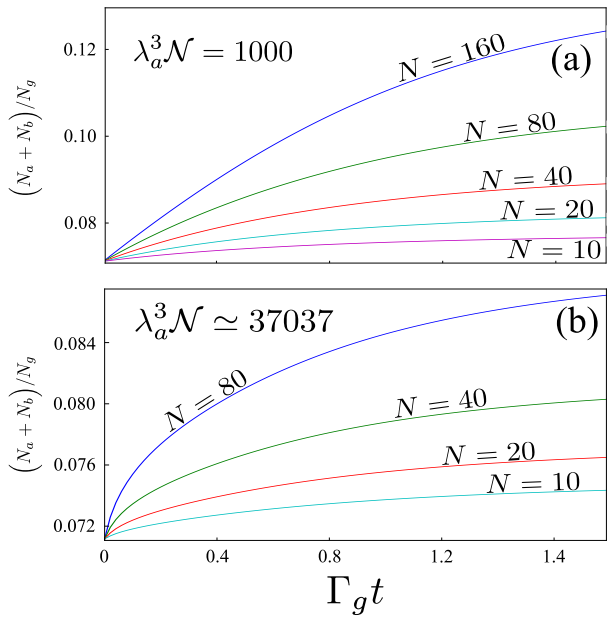}
	\caption{The number of atoms in a cloud that have decayed into either of the two Rydberg states compared to the number of atoms that have decayed into the ground state, $(N_{a} + N_{b})/N_{g}$. This is shown for clouds with various numbers of atoms, $N$,  such that (a) $\lambda_{a}^{3}\mathcal{N} = 1000$ and (b) $\lambda_{a}^{3}\mathcal{N} = 37037$. Note that for every cloud shown, only a very small fraction of the population actually decays into the Rydberg states. For the reasons argued in Sec.~\ref{sec:three}, this diminishes the cooperative behavior of the cloud. This is calculated using Eq.~(\ref{eq:main})}
	\label{fig:v_ground}
\end{figure}

Lastly, one must consider the decay to state $g$, in order to fully understand what is occurring in Fig.~\ref{fig:four}. For reasons argued in Sec.~\ref{sec:three}, the fact that $\Gamma_{g}/\Gamma_{a} \simeq 21$ and $\Gamma_{g}/\Gamma_{b} \simeq 43$ suggests that the decay via the $g$ transition strongly suppresses the cooperative behavior in Fig.~\ref{fig:four}. This is illustrated concretely in Fig.~\ref{fig:v_ground}, where it is shown that for even the most superradiant cloud in Fig.~\ref{fig:four}, less than $15\%$ of the population decays to either of the two Rydberg states. This shows that for the values of $N$ that are currently computationally feasible, the superradiant behavior of a system is dominated by the decay to the ground state. As was argued in Sec~\ref{sec:three}, for clouds with significantly more atoms this effect will be less important.

\section{Conclusion}

This work consisted of the development and implementation of a robust set of differential equations that can be used to study superradiance in ensembles of initially inverted atoms. This set of equations incorporated the dephasing due to elastic dipole-dipole interactions present in dense atomic clouds, and reduced the calculation so that $100$s of multi-level atoms could be simulated. Note that unlike some recent experiments \cite{grimes2016}, the superradiant cascades studied here were not triggered. `Rydberg-like' systems were studied so that two fundamentally different effects could be examined. Sec. \ref{sec:two} focused on dephasing due to the elastic part of the dipole-dipole interaction. Here we found that because the elastic dephasing rate of a given transition is $\propto \lambda^{3}_{\alpha}\mathcal{N}$, where $\mathcal{N}$ is the average density for a given cloud, there is a narrow range of $\lambda_{\alpha}$ such that superradiance can develop significantly. Further in Sec. \ref{sec:three}, it was shown that the presence of an alternate decay channel also suppresses the buildup of superradiance. In Sec. \ref{sec:four}, both of these mechanisms were incorporated into one four-level system. Here the competition between multiple superradiant transitions was studied. Further, it was argued that which transition develops the strongest superradiant behavior depends on the values of $N$, $\Gamma_{\alpha}$ and $\lambda_{\alpha}$ for the system.

The parametrical dependence of a transition's potential to superradiate, described in this work, may provide a starting point for an explanation of why some experiments observe superradiance and some do not. For example, the observations in \cite{walker2008} were conducted for clouds where $\mathcal{N} \sim 10^{7} $~cm$^{-3}$ while in \cite{zhou2016} they were conducted at $\mathcal{N} \sim 10^{9}$~cm$^{-3}$. If elastic dephasing were not present, this two orders of magnitude difference in $\mathcal{N}$ would not have a tremendous effect, since the high lying transitions from an initial Rydberg state are all in the Dicke regime. When elastic dephasing \textit{is} considered, however, the range of $\lambda_{\alpha}\mathcal{N}^{1/3}$ that can superradiate is significantly narrowed, meaning that the specific value of $\mathcal{N}$ for an experiment is extremely important. For transitions from high-lying Rydberg states to nearby Rydberg states, the value of $\Gamma_{\alpha}$ decreases sharply with the difference in principle quantum number, due to the smaller  dipole moment for those transitions. However, $\Gamma_{\alpha}$ for transitions to low-lying states can be two orders of magnitude larger than the Rydberg transitions, due to the fact that $\Gamma_{\alpha}\propto \omega_{\alpha}^{3}$ \cite{gallagher2005}. Therefore it is possible that for very dense clouds, the Rydberg transitions with the largest values of $\Gamma_{\alpha}$ are prevented from superradiating by elastic dephasing (see Sec.~\ref{sec:two}), while decay rates to lower Rydberg states are prevented by the competition between those transitions and the transitions to the ground state, (see Sec. \ref{sec:three}). For more dilute clouds, however, transitions to nearby Rydberg states will undergo significantly less elastic dephasing, and will therefore be much more likely to superradiate. This could be relevant to elucidating the discrepancies currently present in experiments \cite{walker2008,cote2007,zhou2016,grimes2016}.

The diversity of the transitions available in Rydberg atoms implies that they have tremendous potential for studying superradiance. Nevertheless, quantitatively predicting the superradiant decay in a Rydberg cloud is a daunting task. Certain sophisticated approaches to doing this have been conducted \cite{cote2007}; however, these approaches do not incorporate elastic dephasing. This work fills that gap. However, there is clearly a need for further experimental and theoretical developments that provide the community with much needed insights into Rydberg atoms, and superradiance in general.

This material is based upon work supported by the National Science Foundation under Grant No. 1404419-PHY.

\bibliography{bibtex.bib}

%merlin.mbs apsrev4-1.bst 2010-07-25 4.21a (PWD, AO, DPC) hacked
%Control: key (0)
%Control: author (8) initials jnrlst
%Control: editor formatted (1) identically to author
%Control: production of article title (-1) disabled
%Control: page (0) single
%Control: year (1) truncated
%Control: production of eprint (0) enabled
\begin{thebibliography}{45}%
\makeatletter
\providecommand \@ifxundefined [1]{%
 \@ifx{#1\undefined}
}%
\providecommand \@ifnum [1]{%
 \ifnum #1\expandafter \@firstoftwo
 \else \expandafter \@secondoftwo
 \fi
}%
\providecommand \@ifx [1]{%
 \ifx #1\expandafter \@firstoftwo
 \else \expandafter \@secondoftwo
 \fi
}%
\providecommand \natexlab [1]{#1}%
\providecommand \enquote  [1]{``#1''}%
\providecommand \bibnamefont  [1]{#1}%
\providecommand \bibfnamefont [1]{#1}%
\providecommand \citenamefont [1]{#1}%
\providecommand \href@noop [0]{\@secondoftwo}%
\providecommand \href [0]{\begingroup \@sanitize@url \@href}%
\providecommand \@href[1]{\@@startlink{#1}\@@href}%
\providecommand \@@href[1]{\endgroup#1\@@endlink}%
\providecommand \@sanitize@url [0]{\catcode `\\12\catcode `\$12\catcode
  `\&12\catcode `\#12\catcode `\^12\catcode `\_12\catcode `\%12\relax}%
\providecommand \@@startlink[1]{}%
\providecommand \@@endlink[0]{}%
\providecommand \url  [0]{\begingroup\@sanitize@url \@url }%
\providecommand \@url [1]{\endgroup\@href {#1}{\urlprefix }}%
\providecommand \urlprefix  [0]{URL }%
\providecommand \Eprint [0]{\href }%
\providecommand \doibase [0]{http://dx.doi.org/}%
\providecommand \selectlanguage [0]{\@gobble}%
\providecommand \bibinfo  [0]{\@secondoftwo}%
\providecommand \bibfield  [0]{\@secondoftwo}%
\providecommand \translation [1]{[#1]}%
\providecommand \BibitemOpen [0]{}%
\providecommand \bibitemStop [0]{}%
\providecommand \bibitemNoStop [0]{.\EOS\space}%
\providecommand \EOS [0]{\spacefactor3000\relax}%
\providecommand \BibitemShut  [1]{\csname bibitem#1\endcsname}%
\let\auto@bib@innerbib\@empty
%</preamble>
\bibitem [{\citenamefont {Dicke}(1954)}]{dicke1954}%
  \BibitemOpen
  \bibfield  {author} {\bibinfo {author} {\bibfnamefont {R.~H.}\ \bibnamefont
  {Dicke}},\ }\href@noop {} {\bibfield  {journal} {\bibinfo  {journal} {Phys.
  Rev.}\ }\textbf {\bibinfo {volume} {93}},\ \bibinfo {pages} {99} (\bibinfo
  {year} {1954})}\BibitemShut {NoStop}%
\bibitem [{\citenamefont {Bettles}\ \emph {et~al.}(2015)\citenamefont
  {Bettles}, \citenamefont {Gardiner},\ and\ \citenamefont
  {Adams}}]{bettles2015}%
  \BibitemOpen
  \bibfield  {author} {\bibinfo {author} {\bibfnamefont {R.~J.}\ \bibnamefont
  {Bettles}}, \bibinfo {author} {\bibfnamefont {S.~A.}\ \bibnamefont
  {Gardiner}}, \ and\ \bibinfo {author} {\bibfnamefont {C.~S.}\ \bibnamefont
  {Adams}},\ }\href@noop {} {\bibfield  {journal} {\bibinfo  {journal} {Phys.
  Rev. A}\ }\textbf {\bibinfo {volume} {92}},\ \bibinfo {pages} {063822}
  (\bibinfo {year} {2015})}\BibitemShut {NoStop}%
\bibitem [{\citenamefont {Bettles}\ \emph {et~al.}(2016)\citenamefont
  {Bettles}, \citenamefont {Gardiner},\ and\ \citenamefont
  {Adams}}]{bettles2016}%
  \BibitemOpen
  \bibfield  {author} {\bibinfo {author} {\bibfnamefont {R.~J.}\ \bibnamefont
  {Bettles}}, \bibinfo {author} {\bibfnamefont {S.~A.}\ \bibnamefont
  {Gardiner}}, \ and\ \bibinfo {author} {\bibfnamefont {C.~S.}\ \bibnamefont
  {Adams}},\ }\href {\doibase 10.1103/PhysRevLett.116.103602} {\bibfield
  {journal} {\bibinfo  {journal} {Phys. Rev. Lett.}\ }\textbf {\bibinfo
  {volume} {116}},\ \bibinfo {pages} {103602} (\bibinfo {year}
  {2016})}\BibitemShut {NoStop}%
\bibitem [{\citenamefont {Sutherland}\ and\ \citenamefont
  {Robicheaux}(2016{\natexlab{a}})}]{sutherland2016}%
  \BibitemOpen
  \bibfield  {author} {\bibinfo {author} {\bibfnamefont {R.~T.}\ \bibnamefont
  {Sutherland}}\ and\ \bibinfo {author} {\bibfnamefont {F.}~\bibnamefont
  {Robicheaux}},\ }\href {\doibase 10.1103/PhysRevA.93.023407} {\bibfield
  {journal} {\bibinfo  {journal} {Phys. Rev. A}\ }\textbf {\bibinfo {volume}
  {93}},\ \bibinfo {pages} {023407} (\bibinfo {year}
  {2016}{\natexlab{a}})}\BibitemShut {NoStop}%
\bibitem [{\citenamefont {Sutherland}\ and\ \citenamefont
  {Robicheaux}(2016{\natexlab{b}})}]{sutherland2016_2}%
  \BibitemOpen
  \bibfield  {author} {\bibinfo {author} {\bibfnamefont {R.~T.}\ \bibnamefont
  {Sutherland}}\ and\ \bibinfo {author} {\bibfnamefont {F.}~\bibnamefont
  {Robicheaux}},\ }\href {\doibase 10.1103/PhysRevA.94.013847} {\bibfield
  {journal} {\bibinfo  {journal} {Phys. Rev. A}\ }\textbf {\bibinfo {volume}
  {94}},\ \bibinfo {pages} {013847} (\bibinfo {year}
  {2016}{\natexlab{b}})}\BibitemShut {NoStop}%
\bibitem [{\citenamefont {Marek}(1979)}]{marek1979}%
  \BibitemOpen
  \bibfield  {author} {\bibinfo {author} {\bibfnamefont {J.}~\bibnamefont
  {Marek}},\ }\href@noop {} {\bibfield  {journal} {\bibinfo  {journal} {Journal
  of Physics B: Atomic and Molecular Physics}\ }\textbf {\bibinfo {volume}
  {12}},\ \bibinfo {pages} {L229} (\bibinfo {year} {1979})}\BibitemShut
  {NoStop}%
\bibitem [{\citenamefont {Scully}\ \emph {et~al.}(2006)\citenamefont {Scully},
  \citenamefont {Fry}, \citenamefont {Ooi},\ and\ \citenamefont
  {W\'odkiewicz}}]{scully2006}%
  \BibitemOpen
  \bibfield  {author} {\bibinfo {author} {\bibfnamefont {M.~O.}\ \bibnamefont
  {Scully}}, \bibinfo {author} {\bibfnamefont {E.~S.}\ \bibnamefont {Fry}},
  \bibinfo {author} {\bibfnamefont {C.~H.~R.}\ \bibnamefont {Ooi}}, \ and\
  \bibinfo {author} {\bibfnamefont {K.}~\bibnamefont {W\'odkiewicz}},\ }\href
  {\doibase 10.1103/PhysRevLett.96.010501} {\bibfield  {journal} {\bibinfo
  {journal} {Phys. Rev. Lett.}\ }\textbf {\bibinfo {volume} {96}},\ \bibinfo
  {pages} {010501} (\bibinfo {year} {2006})}\BibitemShut {NoStop}%
\bibitem [{\citenamefont {Scully}(2007)}]{scully2007}%
  \BibitemOpen
  \bibfield  {author} {\bibinfo {author} {\bibfnamefont {M.}~\bibnamefont
  {Scully}},\ }\href@noop {} {\bibfield  {journal} {\bibinfo  {journal} {Laser
  Phys.}\ }\textbf {\bibinfo {volume} {17}},\ \bibinfo {pages} {635} (\bibinfo
  {year} {2007})}\BibitemShut {NoStop}%
\bibitem [{\citenamefont {Guerin}\ \emph {et~al.}(2016)\citenamefont {Guerin},
  \citenamefont {Ara\'ujo},\ and\ \citenamefont {Kaiser}}]{kaiser2015}%
  \BibitemOpen
  \bibfield  {author} {\bibinfo {author} {\bibfnamefont {W.}~\bibnamefont
  {Guerin}}, \bibinfo {author} {\bibfnamefont {M.~O.}\ \bibnamefont
  {Ara\'ujo}}, \ and\ \bibinfo {author} {\bibfnamefont {R.}~\bibnamefont
  {Kaiser}},\ }\href {\doibase 10.1103/PhysRevLett.116.083601} {\bibfield
  {journal} {\bibinfo  {journal} {Phys. Rev. Lett.}\ }\textbf {\bibinfo
  {volume} {116}},\ \bibinfo {pages} {083601} (\bibinfo {year}
  {2016})}\BibitemShut {NoStop}%
\bibitem [{\citenamefont {Akkermans}\ \emph {et~al.}(2008)\citenamefont
  {Akkermans}, \citenamefont {Gero},\ and\ \citenamefont
  {Kaiser}}]{kaiser2008_2}%
  \BibitemOpen
  \bibfield  {author} {\bibinfo {author} {\bibfnamefont {E.}~\bibnamefont
  {Akkermans}}, \bibinfo {author} {\bibfnamefont {A.}~\bibnamefont {Gero}}, \
  and\ \bibinfo {author} {\bibfnamefont {R.}~\bibnamefont {Kaiser}},\ }\href
  {\doibase 10.1103/PhysRevLett.101.103602} {\bibfield  {journal} {\bibinfo
  {journal} {Phys. Rev. Lett.}\ }\textbf {\bibinfo {volume} {101}},\ \bibinfo
  {pages} {103602} (\bibinfo {year} {2008})}\BibitemShut {NoStop}%
\bibitem [{\citenamefont {Bromley}\ \emph {et~al.}(2016)\citenamefont
  {Bromley}, \citenamefont {Zhu}, \citenamefont {Bishof}, \citenamefont
  {Zhang}, \citenamefont {Bothwell}, \citenamefont {Schachenmayer},
  \citenamefont {J.T.L.}, \citenamefont {Kaiser}, \citenamefont {Yelin},
  \citenamefont {Lukin}, \citenamefont {Rey},\ and\ \citenamefont
  {Ye}}]{bromley2016}%
  \BibitemOpen
  \bibfield  {author} {\bibinfo {author} {\bibfnamefont {S.}~\bibnamefont
  {Bromley}}, \bibinfo {author} {\bibfnamefont {B.}~\bibnamefont {Zhu}},
  \bibinfo {author} {\bibfnamefont {M.}~\bibnamefont {Bishof}}, \bibinfo
  {author} {\bibfnamefont {X.}~\bibnamefont {Zhang}}, \bibinfo {author}
  {\bibfnamefont {T.}~\bibnamefont {Bothwell}}, \bibinfo {author} {\bibnamefont
  {Schachenmayer}}, \bibinfo {author} {\bibfnamefont {N.}~\bibnamefont
  {J.T.L.}}, \bibinfo {author} {\bibfnamefont {R.}~\bibnamefont {Kaiser}},
  \bibinfo {author} {\bibfnamefont {S.~F.}\ \bibnamefont {Yelin}}, \bibinfo
  {author} {\bibfnamefont {M.}~\bibnamefont {Lukin}}, \bibinfo {author}
  {\bibfnamefont {A.}~\bibnamefont {Rey}}, \ and\ \bibinfo {author}
  {\bibfnamefont {J.}~\bibnamefont {Ye}},\ }\href@noop {} {\bibfield  {journal}
  {\bibinfo  {journal} {Nat. Commun.}\ }\textbf {\bibinfo {volume} {7}}
  (\bibinfo {year} {2016})}\BibitemShut {NoStop}%
\bibitem [{\citenamefont {Day}\ \emph {et~al.}(2008)\citenamefont {Day},
  \citenamefont {Brekke},\ and\ \citenamefont {Walker}}]{walker2008}%
  \BibitemOpen
  \bibfield  {author} {\bibinfo {author} {\bibfnamefont {J.~O.}\ \bibnamefont
  {Day}}, \bibinfo {author} {\bibfnamefont {E.}~\bibnamefont {Brekke}}, \ and\
  \bibinfo {author} {\bibfnamefont {T.~G.}\ \bibnamefont {Walker}},\ }\href
  {\doibase 10.1103/PhysRevA.77.052712} {\bibfield  {journal} {\bibinfo
  {journal} {Phys. Rev. A}\ }\textbf {\bibinfo {volume} {77}},\ \bibinfo
  {pages} {052712} (\bibinfo {year} {2008})}\BibitemShut {NoStop}%
\bibitem [{\citenamefont {Wang}\ \emph {et~al.}(2007)\citenamefont {Wang},
  \citenamefont {Yelin}, \citenamefont {C\^ot\'e}, \citenamefont {Eyler},
  \citenamefont {Farooqi}, \citenamefont {Gould}, \citenamefont
  {Ko\ifmmode~\check{s}\else \v{s}\fi{}trun}, \citenamefont {Tong},\ and\
  \citenamefont {Vrinceanu}}]{cote2007}%
  \BibitemOpen
  \bibfield  {author} {\bibinfo {author} {\bibfnamefont {T.}~\bibnamefont
  {Wang}}, \bibinfo {author} {\bibfnamefont {S.~F.}\ \bibnamefont {Yelin}},
  \bibinfo {author} {\bibfnamefont {R.}~\bibnamefont {C\^ot\'e}}, \bibinfo
  {author} {\bibfnamefont {E.~E.}\ \bibnamefont {Eyler}}, \bibinfo {author}
  {\bibfnamefont {S.~M.}\ \bibnamefont {Farooqi}}, \bibinfo {author}
  {\bibfnamefont {P.~L.}\ \bibnamefont {Gould}}, \bibinfo {author}
  {\bibfnamefont {M.}~\bibnamefont {Ko\ifmmode~\check{s}\else \v{s}\fi{}trun}},
  \bibinfo {author} {\bibfnamefont {D.}~\bibnamefont {Tong}}, \ and\ \bibinfo
  {author} {\bibfnamefont {D.}~\bibnamefont {Vrinceanu}},\ }\href {\doibase
  10.1103/PhysRevA.75.033802} {\bibfield  {journal} {\bibinfo  {journal} {Phys.
  Rev. A}\ }\textbf {\bibinfo {volume} {75}},\ \bibinfo {pages} {033802}
  (\bibinfo {year} {2007})}\BibitemShut {NoStop}%
\bibitem [{\citenamefont {{Grimes}}\ \emph {et~al.}(2016)\citenamefont
  {{Grimes}}, \citenamefont {{Coy}}, \citenamefont {{Barnum}}, \citenamefont
  {{Zhou}}, \citenamefont {{Yelin}},\ and\ \citenamefont
  {{Field}}}]{grimes2016}%
  \BibitemOpen
  \bibfield  {author} {\bibinfo {author} {\bibfnamefont {D.~D.}\ \bibnamefont
  {{Grimes}}}, \bibinfo {author} {\bibfnamefont {S.~L.}\ \bibnamefont {{Coy}}},
  \bibinfo {author} {\bibfnamefont {T.~J.}\ \bibnamefont {{Barnum}}}, \bibinfo
  {author} {\bibfnamefont {Y.}~\bibnamefont {{Zhou}}}, \bibinfo {author}
  {\bibfnamefont {S.~F.}\ \bibnamefont {{Yelin}}}, \ and\ \bibinfo {author}
  {\bibfnamefont {R.~W.}\ \bibnamefont {{Field}}},\ }\href@noop {} {\bibfield
  {journal} {\bibinfo  {journal} {ArXiv e-prints}\ } (\bibinfo {year}
  {2016})},\ \Eprint {http://arxiv.org/abs/1604.03005} {arXiv:1604.03005
  [physics.atom-ph]} \BibitemShut {NoStop}%
\bibitem [{\citenamefont {Roof}\ \emph {et~al.}(2016)\citenamefont {Roof},
  \citenamefont {Kemp}, \citenamefont {Havey},\ and\ \citenamefont
  {Sokolov}}]{roof2016_2}%
  \BibitemOpen
  \bibfield  {author} {\bibinfo {author} {\bibfnamefont {S.~J.}\ \bibnamefont
  {Roof}}, \bibinfo {author} {\bibfnamefont {K.~J.}\ \bibnamefont {Kemp}},
  \bibinfo {author} {\bibfnamefont {M.~D.}\ \bibnamefont {Havey}}, \ and\
  \bibinfo {author} {\bibfnamefont {I.~M.}\ \bibnamefont {Sokolov}},\ }\href
  {\doibase 10.1103/PhysRevLett.117.073003} {\bibfield  {journal} {\bibinfo
  {journal} {Phys. Rev. Lett.}\ }\textbf {\bibinfo {volume} {117}},\ \bibinfo
  {pages} {073003} (\bibinfo {year} {2016})}\BibitemShut {NoStop}%
\bibitem [{\citenamefont {Crubellier}\ \emph {et~al.}(1981)\citenamefont
  {Crubellier}, \citenamefont {Liberman}, \citenamefont {Pillet},\ and\
  \citenamefont {Schweighofer}}]{crubellier1981}%
  \BibitemOpen
  \bibfield  {author} {\bibinfo {author} {\bibfnamefont {A.}~\bibnamefont
  {Crubellier}}, \bibinfo {author} {\bibfnamefont {S.}~\bibnamefont
  {Liberman}}, \bibinfo {author} {\bibfnamefont {P.}~\bibnamefont {Pillet}}, \
  and\ \bibinfo {author} {\bibfnamefont {M.}~\bibnamefont {Schweighofer}},\
  }\href@noop {} {\bibfield  {journal} {\bibinfo  {journal} {Journal of Physics
  B: Atomic and Molecular Physics}\ }\textbf {\bibinfo {volume} {14}},\
  \bibinfo {pages} {L177} (\bibinfo {year} {1981})}\BibitemShut {NoStop}%
\bibitem [{\citenamefont {Scully}(2009)}]{scully2009}%
  \BibitemOpen
  \bibfield  {author} {\bibinfo {author} {\bibfnamefont {M.~O.}\ \bibnamefont
  {Scully}},\ }\href {\doibase 10.1103/PhysRevLett.102.143601} {\bibfield
  {journal} {\bibinfo  {journal} {Phys. Rev. Lett.}\ }\textbf {\bibinfo
  {volume} {102}},\ \bibinfo {pages} {143601} (\bibinfo {year}
  {2009})}\BibitemShut {NoStop}%
\bibitem [{\citenamefont {Scully}(2015)}]{scully2015}%
  \BibitemOpen
  \bibfield  {author} {\bibinfo {author} {\bibfnamefont {M.~O.}\ \bibnamefont
  {Scully}},\ }\href@noop {} {\bibfield  {journal} {\bibinfo  {journal} {Phys.
  Rev. Lett.}\ }\textbf {\bibinfo {volume} {115}},\ \bibinfo {pages} {243602}
  (\bibinfo {year} {2015})}\BibitemShut {NoStop}%
\bibitem [{\citenamefont {Moi}\ \emph {et~al.}(1983)\citenamefont {Moi},
  \citenamefont {Goy}, \citenamefont {Gross}, \citenamefont {Raimond},
  \citenamefont {Fabre},\ and\ \citenamefont {Haroche}}]{moi1983}%
  \BibitemOpen
  \bibfield  {author} {\bibinfo {author} {\bibfnamefont {L.}~\bibnamefont
  {Moi}}, \bibinfo {author} {\bibfnamefont {P.}~\bibnamefont {Goy}}, \bibinfo
  {author} {\bibfnamefont {M.}~\bibnamefont {Gross}}, \bibinfo {author}
  {\bibfnamefont {J.}~\bibnamefont {Raimond}}, \bibinfo {author} {\bibfnamefont
  {C.}~\bibnamefont {Fabre}}, \ and\ \bibinfo {author} {\bibfnamefont
  {S.}~\bibnamefont {Haroche}},\ }\href@noop {} {\bibfield  {journal} {\bibinfo
   {journal} {Physical Review A}\ }\textbf {\bibinfo {volume} {27}},\ \bibinfo
  {pages} {2043} (\bibinfo {year} {1983})}\BibitemShut {NoStop}%
\bibitem [{\citenamefont {Jenkins}\ \emph {et~al.}(2016)\citenamefont
  {Jenkins}, \citenamefont {Ruostekoski}, \citenamefont {Javanainen},
  \citenamefont {Bourgain}, \citenamefont {Jennewein}, \citenamefont
  {Sortais},\ and\ \citenamefont {Browaeys}}]{ruostekoski2016}%
  \BibitemOpen
  \bibfield  {author} {\bibinfo {author} {\bibfnamefont {S.~D.}\ \bibnamefont
  {Jenkins}}, \bibinfo {author} {\bibfnamefont {J.}~\bibnamefont
  {Ruostekoski}}, \bibinfo {author} {\bibfnamefont {J.}~\bibnamefont
  {Javanainen}}, \bibinfo {author} {\bibfnamefont {R.}~\bibnamefont
  {Bourgain}}, \bibinfo {author} {\bibfnamefont {S.}~\bibnamefont {Jennewein}},
  \bibinfo {author} {\bibfnamefont {Y.~R.~P.}\ \bibnamefont {Sortais}}, \ and\
  \bibinfo {author} {\bibfnamefont {A.}~\bibnamefont {Browaeys}},\ }\href
  {\doibase 10.1103/PhysRevLett.116.183601} {\bibfield  {journal} {\bibinfo
  {journal} {Phys. Rev. Lett.}\ }\textbf {\bibinfo {volume} {116}},\ \bibinfo
  {pages} {183601} (\bibinfo {year} {2016})}\BibitemShut {NoStop}%
\bibitem [{\citenamefont {Lee}\ \emph {et~al.}(2016)\citenamefont {Lee},
  \citenamefont {Jenkins},\ and\ \citenamefont {Ruostekoski}}]{lee2016}%
  \BibitemOpen
  \bibfield  {author} {\bibinfo {author} {\bibfnamefont {M.~D.}\ \bibnamefont
  {Lee}}, \bibinfo {author} {\bibfnamefont {S.~D.}\ \bibnamefont {Jenkins}}, \
  and\ \bibinfo {author} {\bibfnamefont {J.}~\bibnamefont {Ruostekoski}},\
  }\href {\doibase 10.1103/PhysRevA.93.063803} {\bibfield  {journal} {\bibinfo
  {journal} {Phys. Rev. A}\ }\textbf {\bibinfo {volume} {93}},\ \bibinfo
  {pages} {063803} (\bibinfo {year} {2016})}\BibitemShut {NoStop}%
\bibitem [{\citenamefont {Jennewein}\ \emph {et~al.}(2016)\citenamefont
  {Jennewein}, \citenamefont {Besbes}, \citenamefont {Schilder}, \citenamefont
  {Jenkins}, \citenamefont {Sauvan}, \citenamefont {Ruostekoski}, \citenamefont
  {Greffet}, \citenamefont {Sortais},\ and\ \citenamefont
  {Browaeys}}]{jennewein2016}%
  \BibitemOpen
  \bibfield  {author} {\bibinfo {author} {\bibfnamefont {S.}~\bibnamefont
  {Jennewein}}, \bibinfo {author} {\bibfnamefont {M.}~\bibnamefont {Besbes}},
  \bibinfo {author} {\bibfnamefont {N.~J.}\ \bibnamefont {Schilder}}, \bibinfo
  {author} {\bibfnamefont {S.~D.}\ \bibnamefont {Jenkins}}, \bibinfo {author}
  {\bibfnamefont {C.}~\bibnamefont {Sauvan}}, \bibinfo {author} {\bibfnamefont
  {J.}~\bibnamefont {Ruostekoski}}, \bibinfo {author} {\bibfnamefont {J.-J.}\
  \bibnamefont {Greffet}}, \bibinfo {author} {\bibfnamefont {Y.~R.~P.}\
  \bibnamefont {Sortais}}, \ and\ \bibinfo {author} {\bibfnamefont
  {A.}~\bibnamefont {Browaeys}},\ }\href {\doibase
  10.1103/PhysRevLett.116.233601} {\bibfield  {journal} {\bibinfo  {journal}
  {Phys. Rev. Lett.}\ }\textbf {\bibinfo {volume} {116}},\ \bibinfo {pages}
  {233601} (\bibinfo {year} {2016})}\BibitemShut {NoStop}%
\bibitem [{\citenamefont {Molina}\ \emph {et~al.}(2016)\citenamefont {Molina},
  \citenamefont {Benito-Mat\'{\i}as}, \citenamefont {Somoza}, \citenamefont
  {Chen},\ and\ \citenamefont {Zhao}}]{yang2016}%
  \BibitemOpen
  \bibfield  {author} {\bibinfo {author} {\bibfnamefont {R.~A.}\ \bibnamefont
  {Molina}}, \bibinfo {author} {\bibfnamefont {E.}~\bibnamefont
  {Benito-Mat\'{\i}as}}, \bibinfo {author} {\bibfnamefont {A.~D.}\ \bibnamefont
  {Somoza}}, \bibinfo {author} {\bibfnamefont {L.}~\bibnamefont {Chen}}, \ and\
  \bibinfo {author} {\bibfnamefont {Y.}~\bibnamefont {Zhao}},\ }\href {\doibase
  10.1103/PhysRevE.93.022414} {\bibfield  {journal} {\bibinfo  {journal} {Phys.
  Rev. E}\ }\textbf {\bibinfo {volume} {93}},\ \bibinfo {pages} {022414}
  (\bibinfo {year} {2016})}\BibitemShut {NoStop}%
\bibitem [{\citenamefont {Monshouwer}\ \emph {et~al.}(1997)\citenamefont
  {Monshouwer}, \citenamefont {Abrahamsson}, \citenamefont {Van~Mourik},\ and\
  \citenamefont {Van~Grondelle}}]{monshouwer1997_2}%
  \BibitemOpen
  \bibfield  {author} {\bibinfo {author} {\bibfnamefont {R.}~\bibnamefont
  {Monshouwer}}, \bibinfo {author} {\bibfnamefont {M.}~\bibnamefont
  {Abrahamsson}}, \bibinfo {author} {\bibfnamefont {F.}~\bibnamefont
  {Van~Mourik}}, \ and\ \bibinfo {author} {\bibfnamefont {R.}~\bibnamefont
  {Van~Grondelle}},\ }\href@noop {} {\bibfield  {journal} {\bibinfo  {journal}
  {The Journal of Physical Chemistry B}\ }\textbf {\bibinfo {volume} {101}},\
  \bibinfo {pages} {7241} (\bibinfo {year} {1997})}\BibitemShut {NoStop}%
\bibitem [{\citenamefont {{Bradac}}\ \emph {et~al.}(2016)\citenamefont
  {{Bradac}}, \citenamefont {{Johnsson}}, \citenamefont {{van Breugel}},
  \citenamefont {{Baragiola}}, \citenamefont {{Martin}}, \citenamefont
  {{Juan}}, \citenamefont {{Brennen}},\ and\ \citenamefont
  {{Volz}}}]{bradac2016}%
  \BibitemOpen
  \bibfield  {author} {\bibinfo {author} {\bibfnamefont {C.}~\bibnamefont
  {{Bradac}}}, \bibinfo {author} {\bibfnamefont {M.}~\bibnamefont
  {{Johnsson}}}, \bibinfo {author} {\bibfnamefont {M.}~\bibnamefont {{van
  Breugel}}}, \bibinfo {author} {\bibfnamefont {B.}~\bibnamefont
  {{Baragiola}}}, \bibinfo {author} {\bibfnamefont {R.}~\bibnamefont
  {{Martin}}}, \bibinfo {author} {\bibfnamefont {M.~L.}\ \bibnamefont
  {{Juan}}}, \bibinfo {author} {\bibfnamefont {G.}~\bibnamefont {{Brennen}}}, \
  and\ \bibinfo {author} {\bibfnamefont {T.}~\bibnamefont {{Volz}}},\
  }\href@noop {} {\bibfield  {journal} {\bibinfo  {journal} {ArXiv e-prints}\ }
  (\bibinfo {year} {2016})},\ \Eprint {http://arxiv.org/abs/1608.03119}
  {arXiv:1608.03119 [quant-ph]} \BibitemShut {NoStop}%
\bibitem [{\citenamefont {Scheibner}\ \emph {et~al.}(2007)\citenamefont
  {Scheibner}, \citenamefont {Schmidt}, \citenamefont {Worschech},
  \citenamefont {Forchel}, \citenamefont {Bacher}, \citenamefont {Passow},\
  and\ \citenamefont {Hommel}}]{scheibner2007}%
  \BibitemOpen
  \bibfield  {author} {\bibinfo {author} {\bibfnamefont {M.}~\bibnamefont
  {Scheibner}}, \bibinfo {author} {\bibfnamefont {T.}~\bibnamefont {Schmidt}},
  \bibinfo {author} {\bibfnamefont {L.}~\bibnamefont {Worschech}}, \bibinfo
  {author} {\bibfnamefont {A.}~\bibnamefont {Forchel}}, \bibinfo {author}
  {\bibfnamefont {G.}~\bibnamefont {Bacher}}, \bibinfo {author} {\bibfnamefont
  {T.}~\bibnamefont {Passow}}, \ and\ \bibinfo {author} {\bibfnamefont
  {D.}~\bibnamefont {Hommel}},\ }\href@noop {} {\bibfield  {journal} {\bibinfo
  {journal} {Nature Physics}\ }\textbf {\bibinfo {volume} {3}},\ \bibinfo
  {pages} {106} (\bibinfo {year} {2007})}\BibitemShut {NoStop}%
\bibitem [{\citenamefont {Baumann}\ \emph {et~al.}(2010)\citenamefont
  {Baumann}, \citenamefont {Guerlin}, \citenamefont {Brennecke},\ and\
  \citenamefont {Esslinger}}]{baumann2010}%
  \BibitemOpen
  \bibfield  {author} {\bibinfo {author} {\bibfnamefont {K.}~\bibnamefont
  {Baumann}}, \bibinfo {author} {\bibfnamefont {C.}~\bibnamefont {Guerlin}},
  \bibinfo {author} {\bibfnamefont {F.}~\bibnamefont {Brennecke}}, \ and\
  \bibinfo {author} {\bibfnamefont {T.}~\bibnamefont {Esslinger}},\ }\href@noop
  {} {\bibfield  {journal} {\bibinfo  {journal} {Nature}\ }\textbf {\bibinfo
  {volume} {464}},\ \bibinfo {pages} {1301} (\bibinfo {year}
  {2010})}\BibitemShut {NoStop}%
\bibitem [{\citenamefont {Friedberg}\ \emph {et~al.}(1972)\citenamefont
  {Friedberg}, \citenamefont {Hartmann},\ and\ \citenamefont
  {Manassah}}]{friedberg1972}%
  \BibitemOpen
  \bibfield  {author} {\bibinfo {author} {\bibfnamefont {R.}~\bibnamefont
  {Friedberg}}, \bibinfo {author} {\bibfnamefont {S.}~\bibnamefont {Hartmann}},
  \ and\ \bibinfo {author} {\bibfnamefont {J.}~\bibnamefont {Manassah}},\
  }\href@noop {} {\bibfield  {journal} {\bibinfo  {journal} {Physics Letters
  A}\ }\textbf {\bibinfo {volume} {40}},\ \bibinfo {pages} {365} (\bibinfo
  {year} {1972})}\BibitemShut {NoStop}%
\bibitem [{\citenamefont {Gross}\ and\ \citenamefont
  {Haroche}(1982)}]{gross1982}%
  \BibitemOpen
  \bibfield  {author} {\bibinfo {author} {\bibfnamefont {M.}~\bibnamefont
  {Gross}}\ and\ \bibinfo {author} {\bibfnamefont {S.}~\bibnamefont
  {Haroche}},\ }\href@noop {} {\bibfield  {journal} {\bibinfo  {journal}
  {Physics Reports}\ }\textbf {\bibinfo {volume} {93}},\ \bibinfo {pages} {301}
  (\bibinfo {year} {1982})}\BibitemShut {NoStop}%
\bibitem [{\citenamefont {Han}\ and\ \citenamefont {Maeda}(2014)}]{han2014}%
  \BibitemOpen
  \bibfield  {author} {\bibinfo {author} {\bibfnamefont {J.}~\bibnamefont
  {Han}}\ and\ \bibinfo {author} {\bibfnamefont {H.}~\bibnamefont {Maeda}},\
  }\href@noop {} {\bibfield  {journal} {\bibinfo  {journal} {Canadian Journal
  of Physics}\ }\textbf {\bibinfo {volume} {92}},\ \bibinfo {pages} {1130}
  (\bibinfo {year} {2014})}\BibitemShut {NoStop}%
\bibitem [{\citenamefont {Zhou}\ \emph {et~al.}(2016)\citenamefont {Zhou},
  \citenamefont {Richards},\ and\ \citenamefont {Jones}}]{zhou2016}%
  \BibitemOpen
  \bibfield  {author} {\bibinfo {author} {\bibfnamefont {T.}~\bibnamefont
  {Zhou}}, \bibinfo {author} {\bibfnamefont {B.~G.}\ \bibnamefont {Richards}},
  \ and\ \bibinfo {author} {\bibfnamefont {R.~R.}\ \bibnamefont {Jones}},\
  }\href {\doibase 10.1103/PhysRevA.93.033407} {\bibfield  {journal} {\bibinfo
  {journal} {Phys. Rev. A}\ }\textbf {\bibinfo {volume} {93}},\ \bibinfo
  {pages} {033407} (\bibinfo {year} {2016})}\BibitemShut {NoStop}%
\bibitem [{\citenamefont {Friedberg}\ \emph {et~al.}(1973)\citenamefont
  {Friedberg}, \citenamefont {Hartmann},\ and\ \citenamefont
  {Manassah}}]{friedberg1973}%
  \BibitemOpen
  \bibfield  {author} {\bibinfo {author} {\bibfnamefont {R.}~\bibnamefont
  {Friedberg}}, \bibinfo {author} {\bibfnamefont {S.~R.}\ \bibnamefont
  {Hartmann}}, \ and\ \bibinfo {author} {\bibfnamefont {J.~T.}\ \bibnamefont
  {Manassah}},\ }\href@noop {} {\bibfield  {journal} {\bibinfo  {journal}
  {Phys. Rep.}\ }\textbf {\bibinfo {volume} {7}},\ \bibinfo {pages} {101}
  (\bibinfo {year} {1973})}\BibitemShut {NoStop}%
\bibitem [{\citenamefont {Carmichael}\ and\ \citenamefont
  {Kim}(2000)}]{carmichael2000}%
  \BibitemOpen
  \bibfield  {author} {\bibinfo {author} {\bibfnamefont {H.}~\bibnamefont
  {Carmichael}}\ and\ \bibinfo {author} {\bibfnamefont {K.}~\bibnamefont
  {Kim}},\ }\href@noop {} {\bibfield  {journal} {\bibinfo  {journal} {Optics
  communications}\ }\textbf {\bibinfo {volume} {179}},\ \bibinfo {pages} {417}
  (\bibinfo {year} {2000})}\BibitemShut {NoStop}%
\bibitem [{\citenamefont {Xu}\ \emph {et~al.}(2013)\citenamefont {Xu},
  \citenamefont {Tieri},\ and\ \citenamefont {Holland}}]{holland2013}%
  \BibitemOpen
  \bibfield  {author} {\bibinfo {author} {\bibfnamefont {M.}~\bibnamefont
  {Xu}}, \bibinfo {author} {\bibfnamefont {D.~A.}\ \bibnamefont {Tieri}}, \
  and\ \bibinfo {author} {\bibfnamefont {M.~J.}\ \bibnamefont {Holland}},\
  }\href {\doibase 10.1103/PhysRevA.87.062101} {\bibfield  {journal} {\bibinfo
  {journal} {Phys. Rev. A}\ }\textbf {\bibinfo {volume} {87}},\ \bibinfo
  {pages} {062101} (\bibinfo {year} {2013})}\BibitemShut {NoStop}%
\bibitem [{\citenamefont {{Hartmann}}(2012)}]{hartmann2012}%
  \BibitemOpen
  \bibfield  {author} {\bibinfo {author} {\bibfnamefont {S.}~\bibnamefont
  {{Hartmann}}},\ }\href@noop {} {\bibfield  {journal} {\bibinfo  {journal}
  {ArXiv e-prints}\ } (\bibinfo {year} {2012})},\ \Eprint
  {http://arxiv.org/abs/1201.1732} {arXiv:1201.1732 [quant-ph]} \BibitemShut
  {NoStop}%
\bibitem [{\citenamefont {Agarwal}(2012)}]{agarwal2012}%
  \BibitemOpen
  \bibfield  {author} {\bibinfo {author} {\bibfnamefont {G.~S.}\ \bibnamefont
  {Agarwal}},\ }\href@noop {} {\emph {\bibinfo {title} {Quantum optics}}}\
  (\bibinfo  {publisher} {Cambridge University Press},\ \bibinfo {year}
  {2012})\BibitemShut {NoStop}%
\bibitem [{\citenamefont {Rehler}\ and\ \citenamefont
  {Eberly}(1971)}]{rehler1971}%
  \BibitemOpen
  \bibfield  {author} {\bibinfo {author} {\bibfnamefont {N.~E.}\ \bibnamefont
  {Rehler}}\ and\ \bibinfo {author} {\bibfnamefont {J.~H.}\ \bibnamefont
  {Eberly}},\ }\href@noop {} {\bibfield  {journal} {\bibinfo  {journal}
  {Physical Review A}\ }\textbf {\bibinfo {volume} {3}},\ \bibinfo {pages}
  {1735} (\bibinfo {year} {1971})}\BibitemShut {NoStop}%
\bibitem [{\citenamefont {Skribanowitz}\ \emph {et~al.}(1973)\citenamefont
  {Skribanowitz}, \citenamefont {Herman}, \citenamefont {MacGillivray},\ and\
  \citenamefont {Feld}}]{feld1973}%
  \BibitemOpen
  \bibfield  {author} {\bibinfo {author} {\bibfnamefont {N.}~\bibnamefont
  {Skribanowitz}}, \bibinfo {author} {\bibfnamefont {I.}~\bibnamefont
  {Herman}}, \bibinfo {author} {\bibfnamefont {J.}~\bibnamefont
  {MacGillivray}}, \ and\ \bibinfo {author} {\bibfnamefont {M.}~\bibnamefont
  {Feld}},\ }\href@noop {} {\bibfield  {journal} {\bibinfo  {journal} {Physical
  Review Letters}\ }\textbf {\bibinfo {volume} {30}},\ \bibinfo {pages} {309}
  (\bibinfo {year} {1973})}\BibitemShut {NoStop}%
\bibitem [{\citenamefont {Damanet}\ and\ \citenamefont
  {Martin}(2016)}]{damanet2016}%
  \BibitemOpen
  \bibfield  {author} {\bibinfo {author} {\bibfnamefont {F.}~\bibnamefont
  {Damanet}}\ and\ \bibinfo {author} {\bibfnamefont {J.}~\bibnamefont
  {Martin}},\ }\href@noop {} {\bibfield  {journal} {\bibinfo  {journal} {arXiv
  preprint arXiv:1606.09372}\ } (\bibinfo {year} {2016})}\BibitemShut {NoStop}%
\bibitem [{\citenamefont {Pellegrino}\ \emph {et~al.}(2014)\citenamefont
  {Pellegrino}, \citenamefont {Bourgain}, \citenamefont {Jennewein},
  \citenamefont {Sortais}, \citenamefont {Browaeys}, \citenamefont {Jenkins},\
  and\ \citenamefont {Ruostekoski}}]{pellegrino2014}%
  \BibitemOpen
  \bibfield  {author} {\bibinfo {author} {\bibfnamefont {J.}~\bibnamefont
  {Pellegrino}}, \bibinfo {author} {\bibfnamefont {R.}~\bibnamefont
  {Bourgain}}, \bibinfo {author} {\bibfnamefont {S.}~\bibnamefont {Jennewein}},
  \bibinfo {author} {\bibfnamefont {Y.~R.}\ \bibnamefont {Sortais}}, \bibinfo
  {author} {\bibfnamefont {A.}~\bibnamefont {Browaeys}}, \bibinfo {author}
  {\bibfnamefont {S.}~\bibnamefont {Jenkins}}, \ and\ \bibinfo {author}
  {\bibfnamefont {J.}~\bibnamefont {Ruostekoski}},\ }\href@noop {} {\bibfield
  {journal} {\bibinfo  {journal} {Phys. Rev. Lett.}\ }\textbf {\bibinfo
  {volume} {113}},\ \bibinfo {pages} {133602} (\bibinfo {year}
  {2014})}\BibitemShut {NoStop}%
\bibitem [{\citenamefont {Goetschy}(2011)}]{goetschy2011}%
  \BibitemOpen
  \bibfield  {author} {\bibinfo {author} {\bibfnamefont {A.}~\bibnamefont
  {Goetschy}},\ }\href@noop {} {\bibfield  {journal} {\bibinfo  {journal}
  {Lumi{\`e}re dans les milieux atomiques d{\'e}sordonn{\'e}s: th{\'e}orie des
  matrices euclidiennes et lasers al{\'e}atoire}\ } (\bibinfo {year}
  {2011})}\BibitemShut {NoStop}%
\bibitem [{\citenamefont {Bienaim\'e}\ \emph {et~al.}(2012)\citenamefont
  {Bienaim\'e}, \citenamefont {Piovella},\ and\ \citenamefont
  {Kaiser}}]{bienaime2012}%
  \BibitemOpen
  \bibfield  {author} {\bibinfo {author} {\bibfnamefont {T.}~\bibnamefont
  {Bienaim\'e}}, \bibinfo {author} {\bibfnamefont {N.}~\bibnamefont
  {Piovella}}, \ and\ \bibinfo {author} {\bibfnamefont {R.}~\bibnamefont
  {Kaiser}},\ }\href {\doibase 10.1103/PhysRevLett.108.123602} {\bibfield
  {journal} {\bibinfo  {journal} {Phys. Rev. Lett.}\ }\textbf {\bibinfo
  {volume} {108}},\ \bibinfo {pages} {123602} (\bibinfo {year}
  {2012})}\BibitemShut {NoStop}%
\bibitem [{\citenamefont {{Facchinetti}}\ \emph {et~al.}(2016)\citenamefont
  {{Facchinetti}}, \citenamefont {{Jenkins}},\ and\ \citenamefont
  {{Ruostekoski}}}]{ruoskekoski2016}%
  \BibitemOpen
  \bibfield  {author} {\bibinfo {author} {\bibfnamefont {G.}~\bibnamefont
  {{Facchinetti}}}, \bibinfo {author} {\bibfnamefont {S.~D.}\ \bibnamefont
  {{Jenkins}}}, \ and\ \bibinfo {author} {\bibfnamefont {J.}~\bibnamefont
  {{Ruostekoski}}},\ }\href@noop {} {\bibfield  {journal} {\bibinfo  {journal}
  {ArXiv e-prints}\ } (\bibinfo {year} {2016})},\ \Eprint
  {http://arxiv.org/abs/1609.08350} {arXiv:1609.08350 [physics.atom-ph]}
  \BibitemShut {NoStop}%
\bibitem [{\citenamefont {Sun}\ and\ \citenamefont
  {Robicheaux}(2008)}]{sun2008}%
  \BibitemOpen
  \bibfield  {author} {\bibinfo {author} {\bibfnamefont {B.}~\bibnamefont
  {Sun}}\ and\ \bibinfo {author} {\bibfnamefont {F.}~\bibnamefont
  {Robicheaux}},\ }\href@noop {} {\bibfield  {journal} {\bibinfo  {journal}
  {Physical Review A}\ }\textbf {\bibinfo {volume} {78}},\ \bibinfo {pages}
  {040701} (\bibinfo {year} {2008})}\BibitemShut {NoStop}%
\bibitem [{\citenamefont {Gallagher}(2005)}]{gallagher2005}%
  \BibitemOpen
  \bibfield  {author} {\bibinfo {author} {\bibfnamefont {T.~F.}\ \bibnamefont
  {Gallagher}},\ }\href@noop {} {\emph {\bibinfo {title} {Rydberg atoms}}},\
  Vol.~\bibinfo {volume} {3}\ (\bibinfo  {publisher} {Cambridge University
  Press},\ \bibinfo {year} {2005})\BibitemShut {NoStop}%
\end{thebibliography}%

\begin{appendices}
Using $g^{\alpha}_{ij} \equiv if^{\alpha}_{ij} + \Gamma^{\alpha}_{ij}/2$, Eq.~(\ref{eq:master}) may be rewritten as:
\begin{eqnarray}\label{eq:simp_master}
&\dot{\hat{\rho}}\nonumber & = \sum_{\alpha}\Gamma_{\alpha}\sum_{j} \Big\{b^{\alpha -}_{j} \hat{\rho} b^{\alpha +}_{j} - \frac{1}{2}b^{\alpha +}_{j}b^{\alpha -}_{j}\hat{\rho} - \frac{1}{2}\hat{\rho}b^{\alpha +}_{j}b^{\alpha -}_{j} \Big\} \\
&+\nonumber & \sum_{\alpha}\sum_{i\neq j}\Big\{ 2Re\big(g^{\alpha}_{ij}\big)b^{\alpha -}_{j}\hat{\rho}b^{\alpha +}_{i}
- g^{\alpha}_{ij} b^{\alpha +}_{i}b^{\alpha -}_{j}\hat{\rho} - g^{\alpha *}_{ij}\hat{\rho}b^{\alpha +}_{i}b^{\alpha -}_{j} \Big\}. \\
\end{eqnarray}
First, we solve for the time-dependence of the operators representing the probablity of atom $i$ being in the state $\alpha$, $\langle b^{\alpha -}_{i}b^{\alpha +}_{i}\rangle$, by substituting this into Eq.~(\ref{eq:one}). Note that for this appendix, we treat the one and two atom terms in Eq.~(\ref{eq:simp_master}) separately, adding them together in the end. Here the single atom terms give:

\begin{eqnarray}
Tr \Big\{ b^{\alpha -}_{m}b^{\alpha +}_{m}\sum_{\beta}\Gamma_{\beta}\sum_{j} (b^{\beta -}_{j}\hat{\rho}b^{\beta +}_{j} - \frac{1}{2}b^{\beta +}_{j}b^{\beta -}_{j}\hat{\rho} - \frac{1}{2}\hat{\rho}b^{\beta +}_{j}b^{\beta -}_{j} )\Big\}. \nonumber \\
\end{eqnarray}
The terms in this sum, such that $m \neq j$ or $\alpha \neq \beta$, will be equal to zero. This gives:
\begin{equation}
\Gamma_{\alpha}\Big\{ \langle b^{\alpha +}_{m} b^{\alpha -}_{m}b^{\alpha +}_{m}b^{\alpha -}_{m} \rangle - \frac{1}{2}\langle b^{\alpha -}_{m}b^{\alpha +}_{m}b^{\alpha +}_{m}b^{\alpha -}_{m} \rangle - \frac{1}{2}\langle b^{\alpha +}_{m}b^{\alpha -}_{m}b^{\alpha -}_{m}b^{\alpha +}_{m}\rangle\Big\}.
\end{equation}
Incorperating the fact that any operator $\propto b^{\alpha +}_{m}b^{\alpha +}_{m}$ or $\propto b^{\alpha -}_{m}b^{\alpha -}_{m}$ is $0$, that $b^{\alpha +}_{m}b^{\alpha -}_{m}b^{\alpha +}_{m}b^{\alpha -}_{m} = b^{\alpha +}_{m}b^{\alpha -}_{m}$, and that $(b^{\alpha +}_{m}b^{\alpha -}_{m}  + \sum_{\beta} b^{\beta -}_{m}b^{\beta +}_{m}= I)$ we obtain:
\begin{eqnarray}
&= \nonumber & \Gamma_{\alpha}\langle b^{\alpha +}_{m}b^{\alpha -}_{m} \rangle \\ &=& \Gamma_{\alpha}\Big\{1 - \sum_{\beta}\langle b^{\beta -}_{m}b^{\beta +}_{m} \rangle \Big\}.
\end{eqnarray}
When considering the two-atom terms, the only non-zero contributions will occur when $\alpha = \beta$, and $m = i$ or $m = j$. This gives:
\begin{eqnarray}
\sum_{j \neq m}&g^{\alpha}_{mj} \nonumber &\langle b^{\alpha -}_{j}b^{\alpha +}_{m} \rangle + \sum_{i\neq m} g^{\alpha *}_{im}\langle b^{\alpha -}_{m}b^{\alpha +}_{i}\rangle \\
&=& 2\sum_{i\neq m}Re\Big\{ g^{\alpha}_{mi}\langle b^{\alpha -}_{i}b^{\alpha +}_{m}\rangle \Big\}.
\end{eqnarray}
Adding these two terms together gives:
\begin{eqnarray}
\frac{d}{dt}\langle b^{\alpha -}_{m}b^{\alpha +}_{m} \rangle &= \nonumber & \Gamma_{\alpha}\Big\{1 - \sum_{\beta}\langle b^{\beta -}_{m}b^{\beta +}_{m}\rangle \Big\} \\ &+& 2\sum_{i\neq m} Re\Big\{ g^{\alpha}_{mi}\langle b^{\alpha -}_{i}b^{\alpha +}_{m}\rangle \Big\}.
\end{eqnarray}

Next we solve for $\frac{d}{dt}\langle b^{\alpha -}_{n}b^{\alpha +}_{m}\rangle$. Pugging $b^{\alpha -}_{n}b^{\alpha +}_{m}$ into the single atom terms in Eq.~(\ref{eq:simp_master}), we find that the only non-zero terms occur when $m = j$ or $n = j$. This gives:
\begin{equation}
-\langle b^{\alpha -}_{n}b^{\alpha +}_{m}\rangle\sum_{\beta}\Gamma_{\beta}.
\end{equation}
Because $\big[ b^{\alpha -(+)}_{m},b^{\alpha -(+)}_{n}\big] = 0$ when $m\neq n$, when considering the two-atom parts of Eq.~(\ref{eq:simp_master}), the only terms that are potentially non-zero occur when:
\begin{center}
\begin{tabular}{ l l l l l}
($i$)  & i = & m & j = & n  \\  
($ii$) & i = & n  & j = & m \\
($iii$) & i = & m & j $\neq$ & n   \\
($iv$) & i $\neq$ & m & j = & n \\
($v$) & i $\neq$ & n  & j = & m \\
($vi$) &i = & n & j $\neq$  & m.
\end{tabular}
\end{center}
Here we calculate each term individually, noting that:
\begin{equation}
 \frac{d}{dt}\langle b^{\alpha -}_{n}b^{\alpha +}_{m}\rangle = (i) + (ii) + (iii) + (iv) + (v) + (vi).
 \end{equation}
\begin{eqnarray}
(i) &= \nonumber & \sum_{\beta}2Re \{g^{\beta}_{mn} \}\Big\{ \langle b^{\beta +}_{m}b^{\alpha +}_{m}b^{\alpha -}_{n}b^{\beta -}_{n}\rangle \\
&- \nonumber & g^{\beta}_{mn}\langle b^{\alpha +}_{m}b^{\alpha -}_{n}b^{\beta +}_{m}b^{\beta -}_{n} \rangle - g^{\beta *}_{mn}\langle b^{\beta +}_{m}b^{\beta -}_{n}b^{\alpha +}_{m} b^{\alpha -}_{n}\rangle \Big\} \\
&=& 0.
\end{eqnarray}

\begin{eqnarray}
(ii) &= \nonumber & \sum_{\beta} 2Re\{ g^{\beta}_{nm}\} \langle b^{\beta +}_{n}b^{\alpha -}_{n}b^{\alpha +}_{m}b^{\beta -}_{m}\rangle \\ &- \nonumber & g^{\beta}_{nm}\langle b^{\alpha -}_{n}b^{\beta +}_{n}b^{\alpha +}_{m}b^{\beta -}_{m} \rangle - g^{\beta *}_{nm} \langle b^{\beta -}_{m}b^{\alpha +}_{m}b^{\beta +}_{n}b^{\alpha -}_{n} \rangle \\  &= \nonumber& 2Re\{ g^{\alpha}_{nm}\} \langle b^{\alpha +}_{n}b^{\alpha -}_{n}b^{\alpha +}_{m}b^{\alpha -}_{m} \rangle \\ &- \nonumber & g^{\alpha}_{nm} \langle b^{\alpha -}_{n}b^{\alpha +}_{n}b^{\alpha +}_{m}b^{\alpha -}_{m}\rangle - g^{\alpha *}_{nm} \langle b^{\alpha -}_{m}b^{\alpha +}_{m}b^{\alpha +}_{n}b^{\alpha -}_{n} \rangle \\ &= \nonumber & 2Re\{ g^{\alpha}_{nm}\} \langle (I - \sum_{\beta} b^{\beta -}_{n}b^{\beta +}_{n})(I - \sum_{\beta}b^{\beta -}_{m}b^{\beta +}_{m}) \rangle \\ &- \nonumber & g^{\alpha}_{nm} \langle b^{\alpha -}_{n}b^{\alpha +}_{n}(I - \sum_{\beta}b^{\beta -}_{m}b^{\beta +}_{m})\rangle \\ &-& g^{\alpha *}_{nm} \langle b^{\alpha -}_{m}b^{\alpha +}_{m}(I - \sum_{\beta}b^{\beta -}_{n}b^{\beta +}_{n}) \rangle
\end{eqnarray}

\begin{eqnarray}
(iii) &= \nonumber & \sum_{\beta}\sum_{j \neq m,n} 2Re\{ g^{\beta}_{mj}\} \langle b^{\beta +}_{m}b^{\alpha +}_{m}b^{\alpha -}_{n}b^{\beta -}_{j}\rangle \\ &- \nonumber & g^{\beta}_{mj}\langle b^{\alpha +}_{m}b^{\alpha -}_{n}b^{\beta +}_{m}b^{\beta -}_{j}\rangle - g^{\beta *}_{mj}\langle b^{\beta +}_{m}b^{\beta -}_{j}b^{\alpha +}_{m}b^{\alpha -}_{n}\rangle \\ &= \nonumber & 0.
\end{eqnarray}

\begin{eqnarray}
(iv) &= \nonumber & \sum_{\beta}\sum_{i\neq m,n} 2Re\{ g^{\beta}_{in} \} \langle b^{\beta +}_{i}b^{\alpha +}_{m}b^{\alpha -}_{n}b^{\beta -}_{n} \rangle \\ &- \nonumber & g^{\beta}_{in}\langle b^{\alpha +}_{m}b^{\alpha -}_{n}b^{\beta +}_{i}b^{\beta -}_{n}\rangle - g^{\beta *}_{in} \langle b^{\beta +}_{i}b^{\beta -}_{n}b^{\alpha +}_{m}b^{\alpha -}_{n}\rangle \\ &=& 0
\end{eqnarray}

\begin{eqnarray}
(v) &= \nonumber & \sum_{\beta}\sum_{i \neq m,n}\Big\{ 2Re\{ g^{\beta}_{im}\}\langle b^{\beta +}_{i}b^{\alpha -}_{n}b^{\alpha +}_{m}b^{\beta -}_{m}\rangle \\ &- \nonumber & g^{\beta}_{im}\langle b^{\alpha +}_{m}b^{\alpha -}_{n}b^{\beta +}_{i}b^{\beta -}_{m}\rangle - g^{\beta *}_{im}\langle b^{\beta +}_{i}b^{\beta -}_{m}b^{\alpha +}_{m}b^{\alpha -}_{n}\rangle \Big\} \\ 
&= \nonumber & \sum_{i \neq m,n} \Big\{ 2Re\{ g^{\alpha}_{im}\}\langle b^{\alpha +}_{i}b^{\alpha -}_{n}b^{\alpha +}_{m}b^{\alpha -}_{m}\rangle \\ &- \nonumber & g^{\alpha}_{im}\langle b^{\alpha +}_{m}b^{\alpha -}_{m}b^{\alpha -}_{n}b^{\alpha +}_{i}\rangle - g^{\alpha *}_{im}\langle b^{\alpha +}_{i}b^{\alpha -}_{n}b^{\alpha -}_{m}b^{\alpha +}_{m}\rangle \\ &- \nonumber & \sum_{\beta \neq \alpha} g^{\beta *}_{im}\langle b^{\beta +}_{i}b^{\alpha -}_{n}b^{\beta -}_{m}b^{\alpha +}_{m}\rangle \Big\} \\
&= \nonumber &\sum_{i\neq m,n} \Big\{ g^{\alpha *}_{im}\langle b^{\alpha +}_{i}b^{\alpha -}_{n}(I - b^{\alpha -}_{m}b^{\alpha +}_{m} - \sum_{\beta}b^{\beta -}_{m}b^{\beta +}_{m})\rangle \\ &-& \sum_{\beta \neq \alpha} g^{\beta *}_{im}\langle b^{\beta +}_{i}b^{\alpha -}_{n}b^{\beta -}_{m}b^{\alpha +}_{m}\rangle \Big\} 
\end{eqnarray}

\begin{eqnarray}
(vi) &= \nonumber & \sum_{\beta}\sum_{j \neq m,n} \Big\{ 2Re\{g^{\beta}_{nj}\}\langle b^{\alpha +}_{m}b^{\beta -}_{j}b^{\beta +}_{n}b^{\alpha -}_{n}\rangle \\ &- \nonumber &g^{\beta}_{nj}\langle b^{\alpha +}_{m}b^{\beta -}_{j}b^{\alpha -}_{n}b^{\beta +}_{n}\rangle - g^{\beta *}_{nj}\langle b^{\beta +}_{n}b^{\alpha -}_{n}b^{\alpha +}_{m}b^{\beta -}_{j}\rangle \Big\} \\ 
&= \nonumber & \sum_{j\neq m,n} \Big\{ 2Re\{g^{\alpha}_{nj}\}\langle b^{\alpha +}_{m}b^{\alpha -}_{j}b^{\alpha +}_{n}b^{\alpha -}_{n}\rangle \\ &- \nonumber & g^{\alpha}_{nj}\langle b^{\alpha +}_{m}b^{\alpha -}_{j}b^{\alpha -}_{n}b^{\alpha +}_{n}\rangle - g^{\alpha *}_{nj}\langle b^{\alpha +}_{n}b^{\alpha -}_{n}b^{\alpha +}_{m}b^{\alpha -}_{j}\rangle \\ &- \nonumber & \sum_{\beta \neq \alpha} g^{\beta}_{nj}\langle b^{\alpha +}_{m}b^{\beta -}_{j}b^{\alpha -}_{n}b^{\beta +}_{n}\rangle \Big\} \\ &= \nonumber & \sum_{j\neq m,n} \Big\{ g^{\alpha}_{nj} \langle b^{\alpha +}_{m}b^{\alpha -}_{j}(I - b^{\alpha -}_{n}b^{\alpha +}_{n} - \sum_{\beta} b^{\beta -}_{n}b^{\beta +}_{n})\rangle \\ &-& \sum_{\beta \neq \alpha} g^{\beta}_{nj}\langle b^{\alpha +}_{m}b^{\beta -}_{j}b^{\alpha -}_{n}b^{\beta +}_{n}\rangle \Big\}
\end{eqnarray}
Adding these all together and rearranging commuting terms gives:
\begin{eqnarray}
&\frac{d}{dt}\nonumber &\langle  b^{\alpha -}_{n} b^{\alpha +}_{m} \rangle = -\langle b^{\alpha -}_{n}b^{\alpha +}_{m} \rangle \sum_{\beta} \Gamma_{\beta} \\ 
&+ \nonumber & \sum_{j\neq m,n} \Big\{ g^{\alpha}_{nj} \langle b^{\alpha -}_{j}b^{\alpha +}_{m}(I - b^{\alpha -}_{n}b^{\alpha +}_{n} - \sum_{\beta} b^{\beta -}_{n}b^{\beta +}_{n})\rangle\Big\} \\
&+ \nonumber &\sum_{j\neq m,n} \Big\{ g^{\alpha *}_{jm}\langle b^{\alpha -}_{n}b^{\alpha +}_{j}(I - b^{\alpha -}_{m}b^{\alpha +}_{m} - \sum_{\beta}b^{\beta -}_{m}b^{\beta +}_{m})\rangle \Big\} \\
 &-\nonumber &\sum_{\beta \neq \alpha} \sum_{j\neq m,n} g^{\beta}_{nj} \langle b^{\alpha -}_{n} b^{\alpha +}_{m} b^{\beta -}_{j} b^{\beta +}_{n} \rangle \\
 &-\nonumber &\sum_{\beta \neq \alpha}\sum_{j \neq m,n} g^{\beta *}_{jm}\langle b^{\beta -}_{m} b^{\beta +}_{j} b^{\alpha -}_{n} b^{\alpha +}_{m} \rangle \\
 &+ \nonumber & 2Re\{ g^{\alpha}_{nm}\} \langle (I - \sum_{\beta} b^{\beta -}_{n}b^{\beta +}_{n})(I - \sum_{\beta}b^{\beta -}_{m}b^{\beta +}_{m}) \rangle \\ &- \nonumber & g^{\alpha}_{nm} \langle b^{\alpha -}_{n}b^{\alpha +}_{n}(I - \sum_{\beta}b^{\beta -}_{m}b^{\beta +}_{m})\rangle \\ &-& g^{\alpha *}_{nm} \langle b^{\alpha -}_{m}b^{\alpha +}_{m}(I - \sum_{\beta}b^{\beta -}_{n}b^{\beta +}_{n}) \rangle.
\end{eqnarray}
This gives a set of differential equations that is not closed. This problem is eschewed by implementing the following factorizations:
\begin{eqnarray}\label{eq:approx}
\langle b^{\alpha -}_{n}b^{\alpha +}_{j}b^{\beta -}_{m}b^{\beta +}_{m} \rangle &\simeq & \langle  b^{\alpha -}_{n}b^{\alpha +}_{j}\rangle \langle  b^{\beta -}_{m} b^{\beta +}_{m} \rangle \nonumber \\
\langle b^{\beta -}_{m}b^{\beta +}_{j} b^{\alpha -}_{n} b^{\alpha +}_{m} \rangle &\simeq & \langle b^{\beta -}_{m}b^{\beta +}_{j}\rangle \langle  b^{\alpha -}_{n}  b^{\alpha +}_{m} \rangle \nonumber \\
 \langle b^{\alpha -}_{m}b^{\alpha +}_{m} b^{\beta -}_{n} b^{\beta +}_{n} \rangle &\simeq & \langle b^{\alpha -}_{m}b^{\alpha +}_{m}\rangle \langle  b^{\beta -}_{n} b^{\beta +}_{n} \rangle \nonumber \\
\langle b^{\alpha -}_{n}b^{\alpha +}_{j}b^{\alpha -}_{m}b^{\alpha +}_{m} \rangle &\simeq & \langle  b^{\alpha -}_{n}b^{\alpha +}_{j}\rangle \langle  b^{\alpha -}_{m} b^{\alpha +}_{m} \rangle \nonumber \\
\langle b^{\alpha -}_{m}b^{\alpha +}_{m} b^{\alpha -}_{n} b^{\alpha +}_{n} \rangle &\simeq & \langle b^{\alpha -}_{m}b^{\alpha +}_{m}\rangle \langle  b^{\alpha -}_{n} b^{\alpha +}_{n} \rangle,
\end{eqnarray}

This results in the final equation given in the text:
\begin{eqnarray}
&\frac{d}{dt}&\langle b^{\alpha -}_{n}b^{\alpha +}_{m} \rangle = - \langle b^{\alpha -}_{n}b^{\alpha +}_{m}  \rangle\sum_{\beta} \Gamma_{\beta} \nonumber \\ 
&+& \sum_{j \neq m,n} g^{\alpha *}_{jm}\langle b^{\alpha -}_{n}b^{\alpha +}_{j} \rangle\Big\{1 - \langle  b^{\alpha -}_{m} b^{\alpha +}_{m} \rangle
 - \sum_{\beta}\langle b^{\beta -}_{m}b^{\beta +}_{m} \rangle \Big\}\nonumber \\ 
  &+& \sum_{j \neq m,n} g^{\alpha}_{nj} \langle  b^{\alpha -}_{j} b^{\alpha +}_{m}\rangle \Big\{ 1 - \langle  b^{\alpha -}_{n} b^{\alpha +}_{n} \rangle
  - \sum_{\beta} \langle b^{\beta -}_{n}b^{\beta +}_{n} \rangle \Big\} \nonumber \\ 
  &-& \sum_{\beta \neq \alpha}\sum_{j \neq m,n}g^{\beta}_{nj}\langle b^{\alpha -}_{n}b^{\alpha +}_{m} \rangle \langle b^{\beta -}_{j}b^{\beta +}_{n} \rangle \nonumber \\ 
 &-& \sum_{\beta \neq \alpha}\sum_{j \neq m,n}g^{\beta *}_{jm}\langle b^{\alpha -}_{n} b^{\alpha +}_{m} \rangle\langle b^{\beta -}_{m} b^{\beta +}_{j}\rangle  \nonumber \\ 
  &+& 2Re\big\{g^{\alpha}_{nm}\big\} (1 - \sum_{\beta}\langle b^{\beta -}_{n}b^{\beta +}_{n}\rangle)(1 - \sum_{\beta}\langle b^{\beta -}_{m}b^{\beta +}_{m}\rangle) \nonumber \\  
  &-& g^{\alpha}_{nm}\langle  b^{\alpha -}_{n} b^{\alpha +}_{n} \rangle (1 - \sum_{\beta}\langle b^{\beta -}_{m}b^{\beta +}_{m}\rangle) \nonumber \\   
  &-& g^{\alpha *}_{nm} \langle b^{\alpha -}_{m}b^{\alpha +}_{m} \rangle(1 - \sum_{\beta}\langle b^{\beta -}_{n}b^{\beta +}_{n}\rangle).
\end{eqnarray}

\end{appendices}
\end{document}